\documentstyle[12pt,aaspp]{article}
\begin{document}
\hfill Preprint

\title{The Dynamics of Flux Tubes in a High {$\beta$} Plasma}

\author{Ethan T. Vishniac}

\affil{Department of Astronomy, University of Texas, Austin, TX 78712,
I: ethan@astro.as.utexas.edu, duncan@astro.as.utexas.edu}

\begin{abstract}
We suggest a new model for the structure of a magnetic field embedded in a
plasma whose average turbulent and magnetic energy densities are both much less
than the gas pressure.  This model is based on the popular notion that the
magnetic field will tend to separate into individual flux tubes.  We point out
that interactions between the flux tubes will be dominated by coherent effects
stemming from the turbulent wakes created as the fluid streams by the flux
tubes. Balancing the attraction caused by shielding effects with turbulent
diffusion we find that flux tubes have typical radii comparable to the local
Mach number squared times the large scale eddy length, are arranged in a one
dimensional fractal pattern, have a radius of curvature comparable to the
largest scale eddies in the turbulence, and have an internal magnetic pressure
comparable to the ambient pressure.  When the average magnetic energy density
is much less than the turbulent energy density the radius, internal
magnetic field,
and curvature scale of the flux tubes will be smaller than these estimates.
Allowing for resistivity changes these properties, but does not alter the
macroscopic properties of the fluid or the large scale magnetic field.  In
either case we show that the Sweet-Parker reconnection rate is much faster than
an eddy turnover time.  Realistic stellar plasmas are expected to either be in
the ideal limit (e.g. the solar photosphere) or the resistive limit (the bulk
of
the solar convection zone).  Allowing for significant viscosity drastically
changes the macroscopic properties of the magnetic field.  We find that all
current numerical simulations of three dimensional MHD turbulence are in the
viscous regime and are inapplicable to stars or accretion disks.  However,
these
simulations are in good quantitative agreement with our model in the viscous
limit.  With the exception of radiation pressure dominated environments, flux
tubes are no more, and often less, buoyant than a diffuse field of comparable
energy density.
\end{abstract}

\section{Introduction}

The study of magnetized plasmas in astrophysics is complicated by
a number of factors, not the least of which is that in considering
magnetic fields in stars or accretion disks, we are considering
plasmas with densities well above those we can study in the laboratory.
In particular, whereas laboratory plasmas are dominated by the
confining magnetic field pressure, stars, and probably accretion
disks, have magnetic fields whose $\beta$ (ratio of gas pressure
to magnetic field pressure) is much greater than one.  Observations
of the Sun suggest that under such circumstances the magnetic
field breaks apart into discrete flux tubes with a small filling
factor.  This trend has also been seen in three dimensional simulations
of MHD turbulence (\cite{nbjrrst92}).  On the other hand, theoretical
treatments of MHD turbulence in high $\beta$ plasmas tend to assume that the
field
is more or less homogeneously distributed throughout the plasma
(e.g. \cite{k65}, and \cite{dc90}).  At the other extreme, there have been
papers (e.g. \cite {do93}) which treat the magnetic field
as a passively advected vector field.  These papers indicate an increasingly
complex substructure, but these calculations are unlikely to be relevant
when considering fields capable of acting back on the surrounding fluid.

Note that although numerical simulations indicate the existence of strong
substructure
(\cite{nbjrrst92}, and \cite{tcv93}),
its exact nature is sensitive to details of the simulation
algorithms and the nature of the large scale flows.  An example of strongly
contrasting results can be found in numerical
simulations by Tao et al. (1993) in which a turbulent flow with
an imposed helicity and a weak diffuse field led to a largely stagnant
and still weak field with substructure, as compared to the
numerical simulations of Hawley \& Balbus (1991) in which a diffuse
field in a shearing flow led to a final state in which the magnetic
pressure was large and continued to drive strong turbulence.

There are at least three reasons for considering the possibility of
substructure in the magnetic field.  First, the mobility of magnetic
field lines in a highly conducting plasma is an important issue,
affecting the dynamics of fluid motion in stars and accretion disks.
Second, the suggestion that turbulent diffusivity does not occur
raises important issues concerning the possibility of creating and
maintaining magnetic fields in astrophysical objects.  For example,
the obvious point that such fields do exist does not ensure that
mean field dynamo theory is a useful tool for describing their
generation.  Third, the possibility that magnetic field lines
tend to concentrate into partially evacuated flux tubes raises
important questions regarding the speed at which such tubes can
rise out of the dynamo region in a star or accretion disk.  If
we assume that an evacuated flux tube of radius $r_t$ is rising
through a medium with a turbulent velocity $V_T$ then equating the
turbulent drag with the buoyant acceleration we have
\begin{equation}
V_b\approx {r_t g \Delta\rho\over V_T \rho},
\end{equation}
where $\Delta\rho/\rho$ is the fractional density depletion of the
flux tube, $V_b$ is the buoyant velocity, and $g$ is the local gravity.
(We have assumed that $V_b\le V_T$ in this expression.)
Clearly we need to know $r_t$ before we can consider the nature of
buoyant magnetic flux loss.

Here we give a qualitative discussion of a simple model for the
distribution of magnetic flux tubes in a turbulent medium.
This paper falls very far short of a derivation of this model
from first principles.  Instead, we simply explore the consequences
of some simple ideas regarding the formation and interaction of
magnetic flux tubes.  We will see that although these ideas cannot
be tested directly in the most interesting regime, i.e. the one
applicable to realistic astrophysical objects, they do yield
quantitative predictions for the current generation of numerical
experiments. In \S 2 we discuss the mechanism by
which small inhomogeneities evolve into discrete flux tubes,
and the size and distribution of such flux tubes.
In \S 3 we allow for the effect of viscosity and resistivity,
both for their intrinsic interest and because no comparison
to numerical results is possible without a quantitative
understanding of their effects.  In \S 4 we discuss reconnection
between the flux tubes and show that it always occurs in
less than an eddy turnover time, even if we calculate the
reconnection rate using the Sweet-Parker rate.
In \S 5 we discuss the implications of this work for
magnetic buoyancy in astrophysical objects.  We find that
our model is consistent with observations of the small
scale structure of the solar magnetic field.  We also
show that magnetic flux loss from accretion disks proceeds
at the same slow rate previously estimated for a diffuse
field, except for radiation pressure dominated disks.
Finally, in \S 6 we conclude with a discussion of some
of the broader issues involved in this work, including
the possibility that the magnetic field fibrils of this
model are an example of a dissipative structure.
In the appendix we compare
this model to numerical simulations of MHD turbulence.

We will see that
there are at least three important consequences
of this model for dynamos and numerical simulations of
dynamos.  First, an initially diffuse
field in a turbulent medium, e.g. a uniform field in a
shearing flow, will initially show exponential growth as the
flux tubes form.  This growth saturates when the flux tube
formation is complete and cannot be used as the basis for a
self-sustaining dynamo effect.  Since the typical state of the
magnetic field is a collection of intense flux tubes, this
effect is of limited interest.  Second,  the organization
of the magnetic field into flux tubes turns out to allow the
field lines to migrate relative to the fluid and to reconnect
efficiently.  In this sense, this model for the magnetic
field substructure implies that the
dynamics of fast dynamos are very much like those of slow dynamos.
Third, this work suggests that the current crop of three
dimensional MHD turbulence simulations are entirely dominated
by viscosity and can be understood in terms of effects which
are negligible in a star or accretion disk.  In other words,
these numerical simulations are inapplicable to realistic
astrophysical objects.

Throughout this paper we take the simplest possible
model for fluid turbulence, i.e. the existence of a stochastic
velocity field with a power spectrum taken from the work of Kolmogorov.
It is likely that intermittency effects will change the details of
the model proposed here.

\section{Magnetic Field Line Distribution in an Ideal Turbulent Fluid}

We begin by considering an idealized situation in which there exists a
turbulent cascade with a well defined large eddy scale
$L_T\equiv 2\pi/k_T$ and a turbulent velocity, on that
scale, of $V_T$.  The fluid is assumed to be inviscid, and perfectly
conducting, although we will also assume that reconnection between magnetic
field lines is efficient.  (We will return to the consistency of these
assumptions later on.)  We will also assume that there is a certain amount of
magnetic flux which crosses a turbulent cell, with an associated
rms Alfv\`en speed $V_A$.  If $V_A\gg V_T$ then the field will
suppress the turbulence.  We will therefore assume that $V_A\le V_T$.
For example, if the magnetic field is in a shearing
flow, surrounded by turbulence of its own creation, then the
near equality of $V_T$ and $V_A$ is guaranteed, as well as the curvature
of the magnetic field lines on the scale $L_T$.

Why should we expect to find flux tubes in a highly conductive fluid?
Normally, one appeals to flux-freezing to establish that matter entrained
on magnetic field lines will remain entrained.  However, this ignores
the possibility that an infinitesimal resistivity can lead to strong collective
effects.  As an example we can consider a flux tube which is thin enough
that it is strongly affected by the motions of the surrounding fluid.
Such a field line will tend to stretch at a rate $\gtrsim k_TV_T$.  If the
plasma
is highly conducting then the same amount of matter will be entrained on
a progressively longer and longer flux tube.  In a stationary state this
stretching will be balanced by the pinching off of closed loops.  These
loops will have some characteristic diameter $l\lesssim L_T$ and a compressive
force
per unit length of $\sim \rho_t l^{-1} V_{At}^2\pi r_t^2$, where $\rho_t$ is
the
density in the tube and $V_{At}$ is the rms Alfv\'en velocity in the tube.
The scale $l$ is determined by the scale on which the flux tube is just weak
enough to bend at large angles in response to velocities on that scale.
This tension will be opposed, usually, by turbulent
stretching with an averaged force per unit length
of $\sim C_d\rho V_T^2 r_t $, which by hypothesis is large
enough to have a significant, but not overwhelming effect.
Some large fraction of the time the loops will collapse (cf. \cite{dfp93})
before they can be reabsorbed by the neighboring flux tubes.
Regardless whether the internal pressure of the
loop is dominated by the magnetic field or gas pressure the magnetic tension
will decrease more slowly than the turbulent stretching force and the loop
will collapse to a plasmoid ball, whose energy is lost either to
microscopic dissipation or the buoyant loss of such magnetic bubbles.
This process will tend to remove matter from the flux tubes
at a rate of $\gtrsim k_TV_T$.  On the other hand, matter will move into
the flux tubes through ohmic diffusion, at a rate
$\sim(\Delta\rho/\rho)\eta/r_t^2$, where $\Delta \rho/\rho$ is the fractional
depletion of matter from the flux tube.  In the limit in which
$\eta\rightarrow 0$ we see that magnetic flux tubes will be perfectly
empty, provided that reconnection isn't suppressed in this limit.
More realistically, how evacuated these flux tubes are will depend on
the efficiency of these loss mechanisms and whether or not mass loading
can take place in the stellar or disk atmosphere. If we start from a uniform,
or
nearly uniform field in an extremely highly conducting fluid,
this process will end when the same amount of flux is
divided into some number of intense flux tubes with a magnetic pressure equal
to the ambient pressure and a local $\beta$ of order unity or less.  The
final rms Alfv\`en velocity will be the geometric mean between its initial
value and the local sound speed.  This initial field
amplification will occur at a rate comparable
to $k_TV_T$, in agreement with the results of numerical experiments
(\cite{hgb94}, \cite{nbjrrst92}).

What will be the typical radius, $r_t$, of the flux tubes? Will they show
correlations for $r>r_t$ or will they be distributed uniformly?
We begin by noting that a flux tube will resist being deformed by turbulent
forces acting on a scale $l$ provided that
\begin{equation}
{B_t^2\over 4\pi R_c}\pi r_t^2>C_d\rho V_l^2 r_t,
\label{eq:stiff}
\end{equation}
where $B_t$ is the magnetic field strength in the flux tube, $C_d$ is
the coefficient of turbulent drag, $V_l$ is the turbulent velocity on
the scale $l$, and $R_c$ is the radius of curvature of the tube.  If
the tube is resisting turbulent motions on a scale $l$ then we can take
$R_c=l/4\equiv k^{-1}\pi/2$.  In addition, if there are bulk forces acting on
both the fluid
and the flux tubes, and exciting the turbulent motions, then the flux tubes
can resist them only if $B_t^2> 4\pi\rho V_T^2$.  As the magnetic field
lines become more and more evacuated,  the ratio of the typical turbulent
force per unit length will scale as $r_t$, whereas the tube stiffness
will scale as $B_t^{3/2} (B_t r_t^2)^{1/2} r_t$, where $B_tr_t^2$ is
proportional
to the magnetic flux is is therefore conserved.  We see that stretching
the flux tubes makes them stiffer as $B_t$ increases, a process that will
continue as long as they can respond effectively to the turbulent motions in
the
fluid. We conclude that they will evolve until they are
relatively straight, in the sense that their
transverse excursions will be small compared to the wavelength of these
disturbances (in the direction of the magnetic field) for all wavelengths
much less than $l$.  Whether or not $l\rightarrow L_T$ will depend on the
presence, or absence, of some dynamo mechanism and the amount of magnetic
flux crossing the boundaries of the system.

What are the forces acting on collection of stiff flux tubes embedded
in a turbulent medium?  First, we note that there is a purely hydrodynamic
attractive force acting between neighboring flux tubes.  Given a bulk
flow with velocity $V_l$ streaming by a flux tube there will be a spreading
turbulent wake, within which the bulk flow will be diminished by roughly
$V_l (r_t/r)^{1/2}$, where $r$ is the distance downstream from the flux tube.
The wake width will be roughly $(r r_t)^{1/2}$.
This implies that a flux tube situated downstream from
its nearest neighbor and possessing a similar large scale curvature
will be subjected to a less intense ram pressure and will
feel a force per volume of $\sim \rho V_l^2 (r_t r)^{-1/2}$
pushing the tube upstream.  However, the full effect of this force
will be felt only by a fraction of order $(r_t/r)^{1/2}$ of the downstream
flux tubes.  Averaged over a loose collection of flux tubes this gives
an upstream force density on the downstream flux tubes of roughly $\sim \rho
V_l^2/r$
per upstream flux tube.
Conversely, given a collection of flux tubes the upstream
tubes will feel more pressure than their downstream companions and
experience a similar average excess bulk force directed downstream.
This is a two dimensional
version of `mock gravity', the attractive force created by shielding
effects in the presence of an isotropic repulsive flux.  In this case
however, the external force is only statistically isotropic.  At any
moment it will have a well defined direction, and the induced attraction
can only act along that axis.

The turbulent wake of a single flux tube will fade into the turbulence
of the fluid if the shear across it is comparable to, or less than,
the shear of the turbulence on the same scale.  Since the strength
of the wake diminishes in proportion to its width, which is proportional
to the square root of its length, this implies that the wake will persist
as long as
\begin{equation}
{V_w \over w}< {V_l\over w}\left({r_t\over w}\right),
\end{equation}
where $w\approx (r r_t)^{1/2}$ is the wake width.
Given $V_w\propto w^{1/3}$, i.e. assuming a Kolmogorov power spectrum,
this condition will be satisfied if
\begin{equation}
r< (r_t l)^{1/2}.
\label{eq:wake}
\end{equation}
In other words, the wake of a single flux tube will persist for a distance
approximately equal to the geometric mean between its width, and the scale
at which it is marginally flexible.
However, if there are other flux tubes within this distance then their
turbulent
wakes will combine and persist to larger scales.

We note that
the normal shearing of the large scale flow will create a dispersive
force density of order $\rho k_T^2V_T^2r$, which is smaller than the
attractive force due to the turbulent wakes.  The shear associated
with small scale eddies, e.g. on a scale $r_t$, is much greater, but there
will be many such eddies along the length of each bundle of flux tubes.
Their effects will add incoherently and their net dispersive effect
will be of order $\rho V_{Tr}^2/L_T\sim V_T^2 (r/L_T)^{2/3}/L_T$, which is
still
negligible.  We conclude that the flux tubes will tend to aggregate,
at least up to the point that the attractive force saturates due to
strong shadowing, i.e. when there are $N$ flux tubes in a region of
size $r$ so that
\begin{equation}
\tau\equiv C_d Nr_t/r\sim 1.
\label{eq:shadow}
\end{equation}
When $\tau$ is below this limit the individual flux tubes feel
a attractive force density towards the center of the bundle of order
\begin{equation}
C_d \tau \rho {V_T^2\over r_t}.
\label{eq:attract}
\end{equation}
This force decreases
when the flux tubes become so stiff that they are essentially straight
regardless of the degree of mutual shadowing (i.e. when the condition
set forth in eq. (\ref{eq:stiff}) is satisfied by a wide margin).

On the other hand, when eq. (\ref{eq:shadow}) is satisfied then it
is unrealistic to treat the interaction of a bundle of semi-rigid flux
tubes with a large scale flow purely in terms of the separate turbulent
wakes created by individual flux tubes.  In particular, in this situation
we can expect to see a collective wake in which the streaming velocity
is reduced by a factor of roughly $1- \tau$ immediately downstream
from the flux tube bundle.  This implies the existence of a Kelvin-Helmholtz
instability with a characteristic growth time of roughly
\begin{equation}
\Gamma_{KH}\approx {V_T\over r}\tau.
\label{eq:kh}
\end{equation}
The velocity associated with the vortices created immediately behind the flux
tube
bundle is roughly
\begin{equation}
V_{vortex}\sim r\Gamma_{KH}\approx V_T\tau.
\label{eq:vkh}
\end{equation}
Since these vortices will be coherent along most of the length of
the flux tube bundle we expect that they will be particularly
important in causing such bundles to disperse.  They will cause
the region immediately downstream from any flux tube to have
significantly larger turbulent pressure than the surrounding flow.
However, the downstream
vortices will tend to be advected away at a velocity at least as
great as the fluid velocity immediately behind the flux tube bundle.
Consequently when $\tau\ll1$ the coherent perturbed velocity near the
flux tube bundle will be less than $V_{vortex}$ by a factor of roughly
$\Gamma_{KH} (r/V_T)$ so that the flux tubes within the bundle will
feel a dispersive bulk force of order
\begin{equation}
C_d\rho \left({V_{vortex}\Gamma{KH}r\over V_T}\right)^2 r_t^{-1},
\end{equation}
or
\begin{equation}
C_d\rho \left(V_T \tau^2\right)^2 r_t^{-1}.
\label{eq:disp}
\end{equation}

By comparing eqs.(\ref{eq:attract}) and (\ref{eq:disp}) we see that
a flux tube bundle in equilibrium will have $\tau\approx 1$ or
$N\approx C_d^{-1} r/r_t$.  Similarly, a collection of $\tilde N$ flux tube
bundles,
each consisting of $N$ flux tubes clustered within a radius of $\tilde r$,
will tend to cluster so as to produce an aggregate
structure with $\tilde N \tilde r/r\sim 1$ or $\tilde N N C_d^{-1}r_t/r\sim 1$.
In other words, the distribution of flux tubes will evolve towards a
fractal of dimension 1, with $N(r)$ flux tubes within a distance $r$
from any given flux tube where
\begin{equation}
N(r)\approx {r\over C_d r_t}.
\label{eq:num1}
\end{equation}
This leads to overlapping turbulent wakes, such that the attractive
force between the flux tubes persists throughout a bundle of
flux tubes.  This fractal distribution of flux tubes persists up the scale
where the turbulent wake of a flux tube bundle extends for a distance
comparable to its size.  We see from eq. (\ref{eq:wake}) that this upper
limit on $r$ is $\sim l$.

The size of an individual flux tube can be derived from the condition
that it be marginally stiff with respect to the surrounding turbulent
motions on the scale $l$, i.e.
\begin{equation}
{B_t^2\over \pi l}\pi r_t^2\approx C_d\rho V_l^2 r_t,
\label{eq:stiff1}
\end{equation}
and that $B_t$ have the maximum sustainable value.
For a perfect fluid the latter
condition is given by pressure equilibrium, i.e.
\begin{equation}
{B_t^2\over 8\pi}=P.
\label{eq:bideal}
\end{equation}
Then the typical radius of a single flux tube is just
\begin{equation}
r_t\approx l{C_d\gamma{\cal M}_l^2\over 8\pi},
\label{eq:radiusa}
\end{equation}
where ${\cal M}_l$ is the Mach number of the turbulent flow on the
scale $l$ and $\gamma$
is its adiabatic index.  Since $l\le L_T$ this gives an upper limit
on the size of the typical flux tube.  To do better we need to invoke
the global properties of the magnetic field.

Summing up the magnetic energy in a collection of flux tubes
we see that if the rms value of the Alfv\'en speed is $V_A$, and
the field consists of $N(l)$ flux tubes per
turbulent cell of size $l$,  then
\begin{equation}
{1\over 2}\rho V_A^2\approx {N(l)w\pi r_t^2P\over l^2},
\label{eq:num}
\end{equation}
and $w$ is a geometric factor describing the amount by which a typical
flux tube has its length increased over $l$ as it crosses the turbulent
cell.  In a saturated state in which the flux tubes are constantly
producing closed loops as the turbulent stresses change this factor will
be of order 3 or 4.  (A flux tube which is almost, but not quite, stretched
enough to produce a closed loop, will have $w$ slightly greater than 2.
Allowing for loops that are not yet pinched off, or have not yet collapsed,
should give a slightly larger value.)
The radius of each of the flux tubes will be given by eq. (\ref{eq:radiusa}),
but a typical flux tube will have a neighbor about that distance away.
Note that this implies very little large scale segregation between the magnetic
field and the turbulent flow.  The magnetic field fills only a small fraction
of the total volume, but the flux tubes are broadly distributed in the fluid.
Of course, this neglects the existence of coherent structures in the flow,
which, if they exist, will tend to collect flux tubes in some fraction of the
total volume.

Eq. (\ref{eq:num1}) implies that on a scale $l$ the number of flux tubes
is approximately $l/(C_d r_t)$.  Combining this result with eqs.
(\ref{eq:num}) and (\ref{eq:radiusa}) we find that
\begin{equation}
V_l^2\approx {4\over w} V_A^2.
\label{eq:equi}
\end{equation}
In other words, the scale $l$ is defined by the condition that the
magnetic field and the turbulent flow be in equipartition on that scale
(any difference between $w$ and 4 being comparable to the cumulative
errors in our estimates of proportionality constants).
If $V_l^2\propto l^n$, where $n=2/3$ for Kolmogorov turbulence, then
we can invert eq. (\ref{eq:equi}) to obtain
\begin{equation}
l\approx L_T\left({4E_B\over wE_T}\right)^{1/n},
\label{eq:scale}
\end{equation}
where $E_B$ and $E_T$ are the magnetic and turbulent energy densities
respectively.  This can be combined with eq. (\ref{eq:radiusa}) to yield
our final answer for the typical flux tube radius in MHD turbulence
in an ideal fluid.  We get
\begin{equation}
r_t\approx L_T{C_d\gamma{\cal M}_T^2\over 8\pi}\left({4E_B\over
wE_T}\right)^{1+1/n}.
\label{eq:radius}
\end{equation}
The number of flux tubes on a scale $l$, which represents the upper limit of
the
fractal clustering pattern, is
\begin{equation}
N(l)={l\over C_d r_t}={8\pi\over C_d^2\gamma {\cal M}_T^2}
\left({wE_T\over 4E_B}\right)^{1+{1\over n}}
\label{eq:numx}
\end{equation}

The total magnetic flux applied across a turbulent cell required to
produce some given $E_B$ depends on the nature of the turbulent flow.  If
we assume the absence of any strong dynamo then the flux on the scale
$L_T$ is related to the flux on the scale $l$ by square root of the number
of independent volumes of length $l$ contained in a two dimensional slice
of a turbulent cell of length $L_T$ then
then
\begin{equation}
\Phi_{tot}\approx {L_T\over l} \Phi_l.
\label{eq:phran}
\end{equation}
Combining this with eqs. (\ref{eq:num1}) and (\ref{eq:radius}) we obtain
\begin{equation}
\Phi_{tot}\approx \left(\sqrt{4\pi\rho} V_A L_T^2\right)
\left[{V_A(\gamma/2)^{1/2}
\over w c_s}\right]\left({4E_B\over wE_T}\right)^{1/n},
\label{eq:flux}
\end{equation}
where the first term in parentheses is the net flux one would expect from a
diffuse field of the same total energy density.  Of course, if there
is a strong dynamo operating in each cell then the total flux might
well be much less, even zero.  On the other hand, in that case we
expect $E_B\approx E_T$ and $l\approx L_T$.  It's interesting to
note that in the limit of an ideal incompressible fluid, the amount of
flux necessary to produce a dynamically significant magnetic field
goes to zero.  It is important to note that in this model the existence
of numerous small fibrils of magnetic field does {\it not} imply
a large number of small scale field reversals.  Given the dynamic
nature of the processes that shape the equilibrium distribution,
and our assumption of rapid reconnection, any complex interweaving
of magnetic field lines pointed in opposing directions will rapidly
relax to a state where the magnetic field direction is the same
for neighboring flux tubes.

We see that even in the
absence of a dynamo an initially uniform field can give rise to
a final state with a greatly amplified magnetic energy density.
It follows that a computer simulation which starts with a uniform field
in a turbulent medium
will see an exponential rise in the magnetic field energy at a rate of
$\gtrsim V_T/L_T$ regardless of whether or not the system in question is
actually capable of supporting a dynamo.
If we define a minimal magnetic energy, $E_i$, by
\begin{equation}
E_i\equiv \left({\Phi_{L_T}\over L_T^2}\right)^2 {1\over 8\pi},
\end{equation}
then from eq. (\ref{eq:flux}) we can show that the system will evolve
to a state where
\begin{equation}
E_B\approx {w\over 4} E_T \left({4PE_i\over E_T^2}\right)^{{n\over 2(1+n)}}.
\end{equation}
Naturally, when this formula gives $E_B>E_T$ it fails, since the imposed
flux gives a flux correlation between volumes of size $l$.  For Kolmogorov
turbulence this implies a net amplification of the magnetic energy equal
to $(E_T/E_i)^{4/5}$ times the Mach number of the turbulence to the
$-0.4$ power.  In a certain sense this is a  `turbulent dynamo', since
statistically symmetric turbulence is driving a large increase in the
magnetic energy, but there is no net production of magnetic flux.  The
increase is a consequence of using a highly artificial initial condition.
It is difficult
to think of realistic circumstances where this effect could be important,
although the very early evolution of the galactic magnetic field might
be one.
If the turbulent energy density is kept equal to the magnetic energy density,
e.g. as in the Velikhov-Chandrasekhar instability (\cite{v59}, \cite{c61}),
the final magnetic
energy density is roughly the geometric mean between $E_i$ and $P$.
(In practice this follows only if the applied flux is in the direction
of the shearing flow (\cite{vd94}) since a field applied in the direction
of the shear vector will drive the creation of a larger magnetic
field in the direction of the flow even in the absence of flux tube
formation.)

Finally, we note that the preceding discussion assumes that the magnetic
flux does not simply migrate to some part of the fluid where the
flow is consistently directed along the field lines.  This will
certainly happen if the velocity field is stationary and well-organized.
Implicit in our assumption of turbulence is that these conditions
are not met and that no such equilibrium is possible.  Consequently,
it is difficult to make any direct comparisons with simulations
of `ABC-flows' (\cite{a65}, \cite{c70}) such as those performed
by Galanti, Sulem, \& Pouquet (1992).

\section{Imperfect Fluids}

In most astrophysical applications one can assume that the viscosity
and magnetic diffusivity are essentially zero.  However, in this case
we will see that in realistic situations, for example most of the
convective region of the Sun, the resistivity of the fluid is large
enough to affect the conclusions of the preceding section.  It is
somewhat more difficult to find cases where viscosity is important,
but it is usually the dominant effect in numerical simulations, which are
our only direct means of testing theories of high $\beta$ MHD turbulence.
In this section we will define the regions of parameter space in which viscous
and diffusive effects dominate.  The boundaries between the various regimes
need to be defined in a four dimensional parameter space since the
Reynolds number, the magnetic Reynolds number, the Mach number of the
turbulence, and the ratio of magnetic to turbulent energies are all
physically significant independent variables.

We start with diffusive effects.  A flux tube of radius $r_t$ and ohmic
diffusivity $\eta$ will spread radially at a rate given by
\begin{equation}
\tau^{-1}_{diff}\sim \eta r_t^{-2},
\end{equation}
where we have taken $\nabla^2 B_t\approx B_t/r_t^2$, which is a reasonable
approximation for a flux tube with a gaussian profile.
This spreading will dominate the evolution of the flux tube if
this rate exceeds the stretching rate for the flux tube (which is
also the large scale shearing rate).  This allows us to define a
diffusive radius of
\begin{equation}
r_{diff}\equiv \left({\eta \over kV_l}\right)^{1/2},
\label{eq:rdiff}
\end{equation}
where $k$ here refers to the wavenumber corresponding to $l$.
If $r_{diff}<r_t$, where $r_t$ is given in eq. (\ref{eq:radius}), then
this defines the skin depth of the flux tube, within which the density
drops sharply from its ambient value.  On the other hand, if $r_{diff}>r_t$,
then typical flux tubes are larger than our previous estimates, with
smaller values of $B_t$.  Combining eqs. (\ref{eq:equi}), (\ref{eq:scale}),
(\ref{eq:radius}) and (\ref{eq:rdiff})
we see that resistivity can be ignored valid if
\begin{equation}
{\cal M}_T^4 \left({V_T\over k_T\eta}\right)\left({4E_B\over w
E_T}\right)^{1/n+5/2}
>\left({4\over C_d\gamma}\right)^2.
\label{eq:diffcrit}
\end{equation}
In other words, the magnetic Reynolds number has to exceed ${\cal M}_T^{-4}$
by about an order of magnitude in order to be safely into the ideal
fluid limit when the magnetic and turbulent energy densities are comparable.
For a Kolmogorov spectrum the ratio of the energies enters into this condition
with an exponent of $4$.  In practice this means that the regime where $E_B\ll
E_T$
is inaccessible to direct simulation, if the results depend on resolving the
smallest flux tubes.  Note that we have defined the Reynolds number here using
the inverse wavenumber.  Another common convention is to use the wavelength
of the turbulence, which gives a larger Reynolds number by a factor of $2\pi$.

When this criterion is not satisfied the fractal distribution of magnetic
flux will be truncated at the scale given by $r_{diff}$.  However, we can still
invoke the notion of marginal resistance to turbulent stretching at larger
scales, so the properties of the system at $r>r_{diff}$ are unchanged.
It is convenient to define a constant $\Psi$ which contains the
environmental parts of the factor
by which a fluid fails to satisfy the criterion set forth in eq.
(\ref{eq:diffcrit}),
i.e.
\begin{equation}
\Psi\equiv \left({\gamma C_d\over 4}\right)^2{\cal M}_T^4
\left({V_T\over k_T\eta}\right).
\label{eq:psidef}
\end{equation}
We conclude that the typical magnetic field intensity in a flux tube is
\begin{equation}
B_t=\sqrt{8\pi P}\Psi^{1/4}\left({4E_B\over wE_T}\right)^{(1/n+5/2)/4}.
\label{eq:btres}
\end{equation}
Consequently, assuming there is no temperature gradient across the
flux tubes, the flux tubes will have a fractional density depletion of
\begin{equation}
{\Delta\rho\over\rho}\approx \Psi^{1/2}\left({4E_B\over
wE_T}\right)^{(1/n+5/2)/2}.
\label{eq:diffrho}
\end{equation}
The total number of flux tubes in a region of size $l$ will
be
\begin{equation}
N={l\over C_d r_t}={2\pi\over C_d}\left({V_l\over k\eta}\right)^{1/2}=
{2\pi\over C_d}\left({V_T\over k_T\eta}\right)^{1/2}\left({4E_B\over
wE_T}\right)^{1/4+1/2n}
\label{eq:numr}
\end{equation}
and the net flux for a turbulent cell of size $L_T$
will be (assuming once again that the flux adds incoherently)
\begin{equation}
\Phi_{tot}\approx \left(\sqrt{4\pi\rho} V_A L_T^2\right)
\left[{V_A(\gamma/2)^{1/2}
\over w c_s}\right]\left({4E_B\over wE_T}\right)^{1/(2n)-5/4}\Psi^{-1/4}.
\label{eq:fluxa}
\end{equation}
Comparing to eq. (\ref{eq:flux}) we see that when
resistivity is important the amount of flux necessary to produce a given
amount of magnetic energy increases relative to the ideal fluid result.

In the regime described above, which we refer to as the
resistive limit, the dynamics of the magnetic field are not strikingly
different just because the smallest flux tubes are no longer as small,
or as completely evacuated.  The only macroscopic field property that changes
is the total magnetic flux associated with a field of a given average energy
density, and it is not clear how significant this quantity is in realistic
situations.   This insensitivity to diffusive effects does not carry over
to the case in which viscous effects
dominate the flux tube dynamics.  This will happen when viscous damping
of the turbulent wakes behind flux tubes prevents the formation of
strong coherent vortices which can spread apart rigid magnetic field lines.
Since the trailing vortices will have radii of about $r_t/2$, and since
the fluid has to complete a full revolution in order to push apart the
field lines, we can approximate the criterion for ignoring viscosity as
\begin{equation}
\left({\pi\over r_t}\right)^2\nu< {V_l\over \pi r_t}.
\label{eq:visc}
\end{equation}
When this inequality is violated the lack of a strongly turbulent wake will
cause flux tubes to aggregate until $r_t$ is large enough to marginally
satisfy this inequality, or until the flux tubes become unresponsive to
the surrounding fluid motions.  Consequently, we can distinguish two regimes
where viscosity has a significant impact and resistivity is negligible.
In the weakly viscous regime
there exists a scale $l<L_T$ such that flux tubes can be marginally
resistant to turbulent motions on that scale with a radius $r_t$ that
just satisfies eq. (\ref{eq:visc}).  In this regime we expect to see
fewer, and larger, flux tubes than we would expect in the ideal fluid
or resistive regimes.  However, flux tubes retain their mobility since
they are still neither completely rigid nor completely entrained in the
surrounding fluid.
In the strongly viscous regime it is possible to form almost
completely rigid flux tubes with typical radii that are small enough that
viscosity can partially damp their turbulent wakes, i.e. that do not
satisfy eq. (\ref{eq:visc}) for $l=L_T$.  In this regime the magnetic field
lines are incapable of undergoing the kind of deformation necessary to
drive a dynamo.  Note that for $E_T\sim E_B$ we expect $l\sim L_T$
in the resistive and ideal fluid regimes.  Consequently, as the viscosity
increases one passes directly into the strongly viscous regime.  The
weakly viscous regime is relevant only when $E_B\ll E_T$.

The boundary between the weakly viscous and ideal fluid regimes can be
derived from eqs. (\ref{eq:radius}) and (\ref{eq:visc}).  We are
in the ideal fluid regime when eq. (\ref{eq:diffcrit}) is satisfied
and
\begin{equation}
1<\chi \left({4E_B\over wE_T}\right)^{3/2+1/n}.
\label{eq:diffvis}
\end{equation}
where
\begin{equation}
\chi\equiv \left({C_d\gamma\over 4\pi^3}\right){\cal M}_T^2
\left({V_T\over k_T\nu}\right).
\label{eq:chidef}
\end{equation}
In other words, the Reynolds number has to exceed ${\cal M}_T^{-2}$ by
roughly two orders of magnitude in order to be in the ideal fluid limit
when $E_B\approx E_T$.  Assuming a Kolmogorov spectrum, this criterion
becomes harder to satisfy for small magnetic energies as the energy
ratio to the third power.
This makes the direct numerical exploration of magnetic field dynamics in the
weak field limit extremely difficult, even setting aside the point that
current simulations do not satisfy this condition for $E_B=E_T$.

In the weakly viscous regime eqs. (\ref{eq:stiff1})
and (\ref{eq:visc}) can be combined to yield
\begin{equation}
l\le L_T \chi^{{-1\over 1+3n/2}},
\label{eq:lwv}
\end{equation}
where the upper limit is obtained by taking $r_t$ as large as possible.
It is reasonable to assume that we are in this limit, provided that
there is more than one flux tube in a volume of size $l$.  If not,
then consolidation of flux tubes will not drive $r_t$ up to the limit
given by eq. (\ref{eq:visc}).  If we are at this limit then
eq. (\ref{eq:lwv}) indicates  that
the scale of curvature for the magnetic field lines is only
a function of the Mach and Reynolds numbers of the turbulent flow.
Combining eqs. (\ref{eq:lwv}), (\ref{eq:visc}) and (\ref{eq:chidef})
we find, after some manipulation, that
\begin{equation}
N(l)={8\pi\over C_d^2\gamma {\cal M}_T^2} \left({4E_B\over wE_T}\right)
\chi^{{2n\over 1+3n/2}}.
\label{eq:numwv}
\end{equation}
However, in this regime the fractal distribution does not extend all
the way from $r_t$ to $l$.  Instead we can show from eqs. (\ref{eq:num1}),
(\ref{eq:visc}), (\ref{eq:chidef}), (\ref{eq:lwv}), and (\ref{eq:numwv})
that it extends to a scale $l_c<l$ given by
\begin{equation}
l_c=l\left({4E_B\over wE_T}\right)\chi^{{n\over 1+3n/2}}.
\end{equation}
In the weakly viscous regime the magnetic field is segregated from the
bulk of the fluid not only on small scales, but also on the scale of
the curvature of the field lines.

When eq. (\ref{eq:numwv}) gives $N(l)\le 1$, then we take $N(l)=1$
and obtain the value of $r_t$ from eqs. (\ref{eq:stiff1}) and
the definition of the energy ratio.  We find
\begin{equation}
r_t={C_d\gamma {\cal M}_T^2\over 4 k_T}\left[\left({4E_B\over w E_T}\right)
{8\pi\over C_d^2\gamma {\cal M}_T^2}\right]^{{1+n\over 2n}},
\end{equation}
and
\begin{equation}
l=L_T\left[\left({4E_B\over w E_T}\right)
{8\pi\over C_d^2\gamma {\cal M}_T^2}\right]^{{1\over 2n}}.
\end{equation}
This weak field extension to the weakly viscous regime obtains when
$N(l)$ in eq. (\ref{eq:numwv}) gives $N(l)<1$ or
\begin{equation}
\chi<\left[\left({4E_B\over w E_T}\right)
{8\pi\over C_d^2\gamma {\cal M}_T^2}\right]^{{-1-3n/2\over 2n}}.
\end{equation}
Its boundary with the strongly viscous limit is defined by $l\rightarrow L_T$
or
\begin{equation}
\left({4E_B\over w E_T}\right)< {C_d^2\gamma {\cal M}_T^2\over 8\pi},
\label{eq:svwv}
\end{equation}
where the inequality is satisfied on the weakly viscous side of the
boundary.

The boundary between the resistive and weakly viscous regimes
is more complicated and involves the appearance of yet another
regime where $\eta$ and $\nu$ are of comparable importance in
determining the typical flux tube radius.  We will refer this
limit as the mixed regime.  From eqs. (\ref{eq:rdiff}) and (\ref{eq:visc})
we see that everywhere in this regime
\begin{equation}
{\eta\over\nu}=\pi^6\left({\nu\over k_TV_T}\right)\left({V_T\over V_l}\right)
\left({L_T\over l}\right).
\label{eq:mixc}
\end{equation}
{}From eq. (\ref{eq:scale}) we see that the boundary between the mixed and
resistive regimes is defined by
\begin{equation}
{\eta\over\nu}<\pi^6\left({\nu\over k_TV_T}\right)
\left({4E_B\over wE_T}\right)^{-1/2-1/n},
\end{equation}
where the inequality is satisfied on the resistive side of the boundary.
Within the mixed regime we have the condition of marginal stability,
eq. (\ref{eq:stiff1}), as well as eq. (\ref{eq:mixc}).  The former
implies that within this regime
\begin{equation}
{B_t^2\over 8\pi P}=\chi\left({l\over L_T}\right)\left({V_l\over V_T}\right)^3.
\label{eq:mixd}
\end{equation}
Combining eqs. (\ref{eq:psidef}), (\ref{eq:chidef}),
(\ref{eq:mixc}) and (\ref{eq:mixd}) we get
\begin{equation}
l=L_T\left({\Psi\over\chi^2}\right)^{{1\over 1+n/2}},
\end{equation}
and
\begin{equation}
{B_t^2\over 8\pi P}=\left({\Psi\over\chi^2}\right)^{{3n/2+1\over1+n/2}}\chi.
\end{equation}
As we move into the mixed regime, in the direction of decreasing
resistivity, the curvature scale of the flux tubes increases, the
radius of the individual tubes decreases (following eq. (\ref{eq:visc}))
and the ratio of magnetic pressure in the flux tubes to the ambient
pressure increases.  Ultimately we either reach the limit
where $l=L_T$ or $B_t^2=8\pi P$.  The former defines the boundary with the
strongly viscous regime.  The condition that we are on the mixed regime
side of the boundary is $\Psi<\chi^2$ or
\begin{equation}
{\eta\over\nu}>\pi^6\left({\nu\over k_TV_T}\right).
\label{eq:svmr}
\end{equation}
The latter limit defines the boundary with the weakly viscous regime, which
can be reached from the mixed regime if, and only if, $\chi>1$.
This boundary is defined by
\begin{equation}
\Psi>\chi^{{1+5n/2\over 1+3n/2}},
\end{equation}
where the inequality is satisfied on the weakly viscous side of the boundary.

Before describing the properties of the strongly viscous regime, it is
useful to state the conditions under which it can be avoided.
{}From eqs. (\ref{eq:lwv}), (\ref{eq:svwv}), and (\ref{eq:svmr})
we see that the boundaries of the strongly viscous regime can
be expressed by the condition that
\begin{equation}
1>\chi^2>\psi,
\end{equation}
and that the ratio of the magnetic energy to turbulent energy exceeds
the Mach number squared divided by something like 25.
In the limit of incompressibility, this implies that eq. (\ref{eq:svmr})
is the only boundary to the strongly viscous regime, and that
satisfying this inequality in a numerical simulation is the
most important goal for numerical simulations of MHD turbulence.
This condition can be expressed as the requirement that the
magnetic Prandtl number has to be less than $\sim 10^{-3}$
times the Reynolds number in order to avoid the strongly viscous
regime.
The difficulty we face in constructing numerical simulations which
will not end up in the strongly viscous regime is
more serious than the failure to reach the ideal fluid regime.
In the strongly viscous regime the magnetic
field will tend to settle into a configuration where the individual flux tubes
are rigid, and yet do not break apart into smaller structures.
Once the field has reached such a configuration it will be largely
insensitive to the surrounding turbulent velocity field.  The rate
at which the field lines stretch to compensate for ohmic diffusion will
be small.  This implies that
such a magnetic field will not grow exponentially due to some net
helicity in the velocity field, even if such growth were expected
from mean-field dynamo theory.  From an astrophysical viewpoint
this is not a particularly interesting limit.  However, it is
the limit most likely to apply to current numerical simulations
of three dimensional MHD turbulence. In the appendix we
present a detailed comparison between various simulations and the
predictions of our model for this limit.

At the edge of the strongly viscous limit the typical flux tube radius is
\begin{equation}
r_t\approx {\pi^3\nu\over V_l}=L_T{C_d\gamma{\cal M}_T^2\over 8\pi}\chi^{-1},
\label{eq:rvisc}
\end{equation}
We note that the fact that viscosity dominates on scales slightly
below  $r_t$
does not prevent us from assuming turbulent drag, although the appropriate
value of $C_d$ will be the value for low Reynolds numbers.

We can then use eq. (\ref{eq:stiff1}) to derive the magnetic field
inside a flux tube.  We find that
\begin{equation}
B_t\approx \sqrt{8\pi P \chi}
\label{eq:btvisc}
\end{equation}
The fractional density depletion will be
\begin{equation}
{\Delta\rho\over\rho}\approx \chi.
\end{equation}
The number of flux tubes per turbulent cell will be
\begin{equation}
N={E_B\over E_T}{\cal M}_T^{-2}\left[{32\pi\over \gamma C_d^2w}\right]\chi,
\label{eq:nvisc}
\end{equation}
where $w$ will be of order 2, since these flux tubes are just thick enough
to bend significantly, but not quite enough to produce loops.
The total magnetic flux will be
\begin{equation}
\Phi_{tot}=\sqrt{4\pi\rho}V_A L_T^2\left({V_A\over c_s}\right)
\left[{\sqrt{\gamma} \over\sqrt{2} w}\right]\chi^{-1/2}.
\end{equation}

In any particular simulation $N$ can be smaller than one.  That is,
it may be that the initial conditions are such that all the flux
in the computational box can be contained in a single flux tube,
regardless of the number of turbulent cells in the box.  However,
it is more likely that given an initially weak diffuse field,
and some local helicity, there will be at least one flux tube per
turbulent cell.  Large scale correlations in the field direction,
which will also be a consequence of some imposed helicity, will
tend to prevent the cancellation of flux tubes in adjacent turbulent
cells. Each flux tube will contain a magnetic flux of
\begin{equation}
\Phi_{tube}\approx \sqrt{8\pi P} L_T^2 \chi^{-3/2} {\cal M}_T^4
{C_d^2\gamma^2\over 64\pi}.
\end{equation}
In this case the magnetic energy density will be
\begin{equation}
E_B\equiv {1\over 2}\rho V_A^2 \approx {1\over 2}\rho V_T^2
\left({\nu k_T\over V_T}\right) {\pi^2\over 8}wC_d.
\label{eq:z7}
\end{equation}
For $\nu$ small this can be arbitrarily far below equipartition,
even if the velocity field of the fluid is capable (under other circumstances)
of driving a strong dynamo.

For moderate Reynolds numbers, i.e. when
\begin{equation}
\left({V_T\over k_T\nu}\right)<\left({2\pi^3\over C_d}\right)
\label{eq:modvisc}
\end{equation}
eq. (\ref{eq:btvisc}) implies $V_{At}<V_T$.
This is misleading since the presence of bulk forces driving
the turbulence can impose $V_{At}\ge V_T$ as a separate condition.
In this case we replace eq. (\ref{eq:btvisc}) with $V_{At}=V_T$.
The condition of marginal stiffness then implies a radius less
than the one given in eq. (\ref{eq:rvisc}), i.e.
\begin{equation}
r_t\approx {L_T C_d\over 4\pi}.
\label{eq:rvisc1}
\end{equation}
The actual radius of a typical flux tube will lie
between this value and the one given in eq. (\ref{eq:rvisc})
depending on the flux threading each turbulent cell.  If
$\Phi_{tot}$ lies in the range
\begin{equation}
\pi\left({C_d\over 2k_T}\right)^2\sqrt{4\pi\rho}V_T<\Phi_{tot}<
\pi\sqrt{4\pi\rho}V_T\left({\pi^3\nu\over V_T}\right)^2
\label{eq:zz}
\end{equation}
there will be one flux tube per turbulent cell with a
radius
\begin{equation}
r_t\approx\left({\Phi_{tot}\over\pi\sqrt{4\pi\rho}V_T}\right)^{1/2}.
\end{equation}
For fluxes beyond the upper limit in eq. (\ref{eq:zz})
the number of flux tubes per turbulent cell is
\begin{equation}
N\approx {V_A^2L_T^2\over\pi^7\nu^2w},
\end{equation}
and the total flux is
\begin{equation}
\Phi_{tot}=\sqrt{4\pi\rho}V_A L_T^2\left({V_A\over V_T}\right)
w^{-1}.
\end{equation}

In this limit the minimal stationary state for a numerical simulation
with an initially weak diffuse field  has a magnetic energy density of
\begin{equation}
{1\over 2}\rho V_A^2\approx {1\over 2}\rho V_T^2 {wC_d^2\over 16\pi}.
\label{eq:min}
\end{equation}
It may seem surprising that this is insensitive to the value of $\nu$,
but this is somewhat misleading.  For $\nu$ so large that the flux tube
wakes become entirely dominated
by viscosity the appropriate value of $C_d$ will scale as $\nu^{1/2}$.
Eq. (\ref{eq:min}) will only apply over a limited range of moderate
Reynolds numbers.
At low Reynolds numbers, when $V_{At}\sim V_T$ and eq.
(\ref{eq:rvisc1}) is valid, resistivity is negligible provided that
\begin{equation}
{V_T\over k_T\eta}>{4\over C_d^2}.
\label{eq:q1}
\end{equation}

There is one other limit on the magnetic Prandtl number which is
important.  If the resistivity, $\eta$, is much larger than
$max[\nu, C_d r_t V_T]$, then the magnetic
flux tubes will tend to resist deforming in response to turbulent
fluid motions (cf. \cite{b50}).  In the ideal fluid limit we
can see from eq. (\ref{eq:radius}) that this implies
\begin{equation}
\eta< {C_d^2\gamma\over 4k_T} V_T{\cal M}_T^2.
\end{equation}
Using the definition of $\Psi$ in eq. (\ref{eq:psidef}) we can
rephrase this as
\begin{equation}
\Psi>{\gamma\over 4}{\cal M}_T^2,
\end{equation}
which is always satisfied since, by definition, a system in the
ideal fluid regime will have $\Psi>1$.
In the resistive regime we need to replace eq. (\ref{eq:radius}) with
eq.(\ref{eq:rdiff}) so that our limit on $\eta$ becomes
\begin{equation}
\eta<C_d^2{V_T\over k_T},
\end{equation}
i.e. the magnetic Reynolds number has to be greater than $C_d^{-2}$,
which is of order unity.  Again, this will be trivially satisfied
in any case of interest.  On the other hand, if we turn our attention
to the viscous regime we see that for high Reynolds numbers
(when the flux tube radius is given by eq. (\ref{eq:rvisc})),
the upper limit on $\eta$ becomes
\begin{equation}
\eta<C_d \pi^3\nu.
\label{eq:evlim}
\end{equation}
At moderate Reynolds numbers the minimum value of $r_t$ is
given by eq. (\ref{eq:rvisc1}) and the limit on $\eta$
becomes
\begin{equation}
\eta<{C_d^2\over 2k_T}V_T.
\label{eq:evlim1}
\end{equation}
Of course, at very small Reynolds numbers we recover the
condition $\eta<\nu$.  The original work by Batchelor proposed
this as the only limit, but based on a very different conceptual
model for the distribution of magnetic flux.  For our purposes
this limit is important only for very small Reynolds numbers.  The less
stringent limits given in eqs. (\ref{eq:evlim}) and (\ref{eq:evlim1})
are the important ones.

In this context, it is interesting to note that
simulations done with $\eta\ge\nu$
have shown a strong suppression of the growth of the
magnetic field energy (\cite{nbjrrst92}).  In fact, these simulations
seem to show that in the viscous regime the critical value of $\eta/\nu$
is close to one (although no such suppression was seen in the work
of Tao et al. (1992) which had $\eta=\nu$).  Nordlund et al. ascribed
the difference to the different heat conductivities used.
Here we suggest that it is instead by due to the different boundary
conditions used in the two simulations.  Tao et al. started with
a weak nonzero flux and a static fluid which gradually responded to
the large scale forcing.  Nordlund et al. used a weak flux with large
scale structure which averaged to zero over the simulation volume, and
began from fully developed turbulence.  Both simulations started from a
uniform field.  Apparently these differences made the simulation
of Nordlund et al. more vulnerable to immediate turbulent dissipation.
The argument in
the preceding paragraph suggests that Nordlund et al. would have
seen little effect if they had taken the final state of a low $\eta$
run and used it as the initial condition for a high $\eta$ run
(assuming that $\eta$ still satisfies the limit given in eq.
(\ref{eq:evlim})).

These results suggest that the optimum strategy for designing a code
which can simulate 3D MHD turbulence in the resistive regime is
to take the largest resistivity consistent with some reasonable
ability to resolve flux tubes. Since the ratio of flux tube radius
to eddy size is roughly the square root of the magnetic Reynolds number
this implies a magnetic Reynolds number of $\sim10^2$.  Fixing this
and maximizing the Reynolds number should give the easiest route
out of the viscous regime.

\section{Reconnection}

In the preceding section we have assumed that reconnection is rapid,
in the sense that flux tubes reconnect much faster than an eddy
turnover time.  In fact, this is a controversial point.  The
actual rate of reconnection depends on the structure of the flux
tubes.  Even in the context of a particular model for their structure
the rate is not well understood.  Parker (1957) and Sweet (1958)
proposed that reconnection should cross a flux tube at a speed of
\begin{equation}
V_{rec}\approx V_A \left({V_A r_t\over \eta}\right)^{-1/2}.
\label{eq:r1}
\end{equation}
The physical basis for this estimate is that at the interface
between two reconnecting flux tubes the gas builds up an excess
pressure, of order $\rho V_A^2$, preventing the opposing field lines from
reconnecting
efficiently.
The rate of reconnection is then controlled by the rate at which
particles escape from the reconnection region, presumably by moving
a distance of order $r_t$ along the field lines.  The excess
pressure is maintained through the heat released by the dissipation
of magnetic field energy.  If we identify $V_A$ with its rms
value, or with the local turbulent velocity, then this rate is
quite slow.  Magnetized regions on scales comparable to
the size of a convective cell are unable to reconnect efficiently
in one eddy turnover time.  This has led to a number of proposals
for mechanisms that will increase reconnection rates.
Several of these (\cite {ce77}, \cite {d84}, and \cite {s88})
appeal to plasma effects which will be heavily
suppressed in a strongly collisional plasma, like the kind we are
considering here.  One process that can apply in
a collisionally dominated plasma is the Petscheck
reconnection mechanism (\cite{p64}) which has the effect of replacing
the denominator
of eq. (\ref{eq:r1}) with the logarithm of the magnetic Reynolds
number.  It remains unclear whether or not this rate is attainable
under realistic conditions inside a star or accretion disk
(e.g. see \cite{b86}).  Another possibility is that once
reconnection gets under way the rate is determined by nonlinear
hydrodynamic processes that increase the reconnection speed by some large
factor (\cite{ml86}).

Here we will assume that the Sweet-Parker rate
is essentially correct.  Given that our purpose is to show that reconnection
is rapid in the model for MHD turbulence we propose here,
this is the conservative strategy.  We will also assume that the
internal structure of a typical flux tube is given by balancing the
effects of stretching along the field lines with radial ohmic diffusion.
In other words, we neglect turbulent diffusion.  We will defer
discussion of this point to a later paper.  Here we note
only that the large magnetic field strength in the flux tubes
may will the dimensionality of the flow.  Since turbulent
diffusion is largely suppressed in two dimensions (\cite{cv91})
it seems plausible to neglect it here (\cite{d94}).
If we model the flux tube as having infinite extent in the $\hat z$
direction, with $\vec B=B(r) \hat z$ and $\partial_z v_z=\tau^{-1}$
then the stationary solution for a tube in an isothermal fluid satisfies
\begin{equation}
{1\over r}\partial_r (rv_r \rho)+{\rho\over \tau}=0,
\label{eq:cont}
\end{equation}
\begin{equation}
{1\over r}\partial_r (rv_r B)={1\over r}\partial_r(r\eta\partial_r B),
\label{eq:bdiff}
\end{equation}
and
\begin{equation}
{B^2\over 8\pi}+\rho {k_B T\over \mu}=P_{tot},
\label{eq:press}
\end{equation}
where $\mu$ is the mean mass per particle, $P_{tot}$ is the total pressure
in the fluid, $\eta$ is the resistivity (assumed to be independent of
density), and $v_r$ is the radial velocity.  Eq. (\ref{eq:bdiff}) can be
integrated, assuming that the magnetic field and its derivative vanish
at large radii, to obtain
\begin{equation}
v_r=\eta\partial_r \ln B={\eta\over 2} \partial_r \ln P_{mag},
\label{eq:vdiff}
\end{equation}
where $P_{mag}$ is the magnetic pressure.
Combining eqs. (\ref{eq:cont}), (\ref{eq:press}) and (\ref{eq:vdiff}) we find
that
\begin{equation}
\left[\partial_x\ln(x\rho)\right]\left[\partial_x\ln(1-\rho/\rho_{\infty})
\right]
+\partial_x^2\ln(1-\rho/\rho_{\infty})+2=0,
\label{eq:flux1}
\end{equation}
where $\rho_{\infty}$ is the density at large distances from the flux tube,
and $x$ is the dimensionless radial distance defined by
\begin{equation}
x\equiv {r\over \sqrt{\eta \tau}}.
\end{equation}
Eq. (\ref{eq:flux1}) can be rewritten in terms of $P_{mag}$ as
\begin{equation}
\left[\partial_x\ln(x(1-P_{mag}/P_{tot}))\right]\left[\partial_x\ln
P_{mag}\right]
+\partial_x^2\ln P_{mag}+2=0.
\label{eq:flux2}
\end{equation}

A flux tube in the resistive regime will have $P_{mag}\ll P_{tot}$ everywhere.
In this case eq. (\ref{eq:flux2}) implies
\begin{equation}
P_{mag}\approx P_{mag}(x=0)\exp \left[-{r^2\over 2\eta\tau}\right].
\end{equation}
A flux tube in the ideal fluid regime has the curious feature that
it consists, in this approximation, of a thin shell with a radius comparable
to $\sqrt{\eta\tau}$ surrounding an interior where $\rho=0$.
If the radius of the
tube is much greater than the thickness of the shell then the full
equation for $\rho$ reduces to an integral solution of the form
\begin{equation}
x-x_0=\int_0^{\rho/\rho_{\infty}} {qdq\over 2(1-q)\sqrt{-\ln(1-q)-q-q^2/2}},
\end{equation}
where $x_0$ is the dimensionless radius of the evacuated interior.
Close to $x_0$ this becomes
\begin{equation}
\rho={\rho_{\infty}\over 3}(x-x_0)^2
\left(1-{5\over 36}(x-x_0)^2+\cdots\right).
\end{equation}
At large distances we find
\begin{equation}
\rho\approx \rho_{\infty}\left(1-\exp[ -(x-x_0-1.45)^2]\right).
\end{equation}

Now we consider reconnection involving two isolated flux tubes
in the ideal fluid limit.  Since the magnetic pressure in the tubes
equals the ambient pressure the Alfv\'en velocity at the
edge of the flux tubes will be roughly $c_s$.  However, subject to
the approximation that $\eta$ is really independent of density and
that the flux tube reaches a stationary internal state,  the
Alfv\'en velocity goes to infinity within a distance of
$r_{skin}\sim \sqrt{\eta\tau}$.  Realistically it will reach a value
exponentially higher than $c_s$.
Consequently, the reconnection rate is just
\begin{equation}
{V_{rec}\over r_{skin}}\approx \left({c_s \eta\over r_t
r_{skin}^2}\right)^{1/2},
\label{eq:rec1b}
\end{equation}
or using eqs. (\ref{eq:scale}) and (\ref{eq:radius})
\begin{equation}
{V_{rec}\over r_{skin}}\approx k_Tc_s{\cal M}_T^{-1/2}
\left({4E_B\over wE_T}\right)^{-{1\over4}-{1\over n}}.
\label{eq:rec1a}
\end{equation}
We note that this estimate is insensitive to $\eta$.

One loophole in this argument is the assumption that reconnection in the outer
layers of the flux tube, where the magnetic field is very small, has a
negligible
effect on the overall speed of reconnection.   Since $V_{rec}\rightarrow 0$
exponentially in this region it is easy to imagine that the actual rate of
reconnection is controlled in the flux tube envelope.  This is the case,
but fortunately this doesn't change our basic conclusion that reconnection
is rapid.  Neglecting reconnection for the moment, we can consider the
dynamics of the collision between two flux tubes.  As the flux tubes
try to move past one another at a speed $\sim v_l$, they create sharp
bends over some region
of typical size $r_t$.  At the contact point the local pressure rises by
roughly $\bar B^2 sin\phi$, where $\phi$ is the bending angle and $\bar B$
is the root mean square value of the magnetic field in the flux tube.
This implies that eq. (\ref{eq:rec1a}) should be corrected by a factor
of $\sim \phi^{1/4}$.  In the ideal fluid limit the flux tube core is
empty, and in that region $V_A=\infty$ (neglecting a breakdown
of our assumption that $\eta$ is independent of density).  However,
bending waves in a flux tube will involve moving the mass contained
in the flux tube skin and so the effective Alfv\'en speed, which is
also the speed at which bending waves can travel down the flux tube,
will be approximately
\begin{equation}
V_{At}\approx \left({2 r_{skin}\over r_t}\right)^{-1/2}c_s
\sim \Psi^{1/4}c_s
\left({4E_B\over wE_T}\right)^{{1\over4n}+{5\over 8}}.
\label{eq:speed}
\end{equation}
When $k V_{At}$ exceeds the reconnection rate, then the Alfv\'en speed
is effectively infinite and the bending angle $\phi$ is just $k V_l t$.
Equating the time in this expression with the characteristic reconnection
time we find that in this limit the reconnection rate given in
eq. (\ref{eq:rec1a}) is modified to
\begin{equation}
{V_{rec}\over r_{skin}}\approx k_Tc_s{\cal M}_T^{-{1\over 5}}
\left({4E_B\over wE_T}\right)^{-{1\over10}-{1\over n}}.
\label{eq:rec1aa}
\end{equation}

Large scale reconnection events will involve bundles of such flux tubes,
each consisting of $N(l)$ individual flux tubes in a bundle of radius $l$.
These tubes do not need to reconnect serially, but simultaneous
reconnection is also unlikely.
The total reconnection rate should be down from the single reconnection
rate given in eq. (\ref{eq:rec1aa}) by at most a factor of $N^{1/2}$.
If the rate of reconnection is slow enough that tubes undergo some
compression before the reconnection front reaches them then this can be
an underestimate of the true reconnection rate.  In no case
can the reconnection rate exceed the bulk flow rate across a bundle
radius.  Using eqs. (\ref{eq:numx}), and (\ref{eq:rec1aa})
we conclude that the bundle reconnection rate divided by the eddy
turnover rate on the scale of curvature of the flux tubes will be
\begin{equation}
{1\over kV_l\tau_{rec}}= \min\left[{\cal M}_T^{-1/5}
\left({4E_B\over wE_T}\right)^{-{1\over 10}},1\right],
\label{eq:reca1}
\end{equation}
where the minimum of 1 for this ratio arises from the fact that the
bundles cannot reconnect in less time than it takes for the flux tubes
to move across a distance $l$.  This minimum rate may be overly
conservative,
if the conditions that lead to flux bundle collisions tend to concentrate
them in the process, or if the process of reconnection itself results
in bulk motions that accelerate the collision.  Ignoring this point
we see that
in this limit of the ideal fluid regime, where the mass loading on the
flux tubes is insignificant, the limiting rate in the
reconnection of flux tube bundles is set by the bulk motion of the
bundles, not by the actual reconnection of the flux tubes.

When the distance a signal can travel down a flux tube in the time for
a pair of isolated flux tubes to reconnect is less than
$l$, the bending angle becomes $V_l/V_{At}$.  From eqs. (\ref{eq:speed})
and (\ref{eq:rec1aa}) this happens when
\begin{equation}
\Psi<{\cal M}_T^{-4/5}\left({4E_B\over wE_T}\right)^{-2.9-{1\over n}}.
\end{equation}
In this case eq. (\ref{eq:rec1a}) needs to be corrected by multiplying
the LHS by $(V_l/V_{At})^{1/4}$.  This gives
\begin{equation}
{V_{rec}\over r_{skin}}\approx k_Tc_s{\cal M}_T^{-{1\over 4}}\Psi^{-1/16}
\left({4E_B\over wE_T}\right)^{-{9\over32}-{17\over 16n}}.
\label{eq:rec1ab}
\end{equation}
We see that the reconnection rate increases slightly as we approach the
boundary of the ideal fluid regime.  In this regime the bundle
reconnection rate is
\begin{equation}
{1\over kV_l\tau_{rec}}= \min\left[{\cal M}_T^{-1/4}\Psi^{-1/16}
\left({4E_B\over wE_T}\right)^{-{9\over 32}-{1\over16n}},1\right].
\label{eq:reca2}
\end{equation}
At the limit of the ideal fluid regime, given by eq. (\ref{eq:diffcrit}),
the bundle reconnection rate estimate obtained by summing up individual
reconnection events can be as large as $kV_l {\cal M}_T^{-1/4}$ times
the energy ratio to the $-1/8$ power.

In the resistive limit $r_{skin}=r_t=\sqrt{\eta\tau}$.  Although
the flux tubes are solidly filled in, the magnetic field within the
tubes is well below equipartition with the exterior pressure.
Consequently $V_{At}$ is still given by eq. (\ref{eq:speed})
and the reconnection rate for a pair of isolated flux tubes
becomes
\begin{equation}
{V_{rec}\over r_{skin}}\approx k_Tc_s{\cal M}_T^{-{1\over 4}}\Psi^{5/16}
\left({4E_B\over wE_T}\right)^{{21\over32}-{11\over 16n}}.
\label{eq:rec1ac}
\end{equation}
Note that now the reconnection rate decreases as the resistivity increases.
Using eq. (\ref{eq:numr}) we see that the bundle reconnection rate becomes
\begin{equation}
{1\over kV_l\tau_{rec}}= \min\left[{\cal M}_T^{-1/4}\Psi^{1/16}
\left({4E_B\over wE_T}\right)^{-{1\over 32}-{1\over16n}},1\right].
\label{eq:reca3}
\end{equation}
We note that ${\cal M}_T^{-1/4}\Psi^{1/16}$ is more or less the sixteenth
root of the the magnetic Reynolds number.  We see that in the resistive
regime the reconnection rate for the individual tubes in a bundle
is slow enough that the eddy turnover rate wins by only a modest factor.
Nevertheless, it does win.

We conclude that assuming the Sweet-Parker rate for magnetic reconnection
in a turbulent medium gives reconnection which happens faster than the
eddy turnover time, so that magnetic reconnection is primarily limited by
the rate at which flux tubes move across the fluid.
Paradoxically, the tube reconnection rate actually slows as the resistivity
in the fluid increases.

\section{Buoyancy}

How quickly will a single flux tube rise?  Each flux tube will feel a bulk
upward acceleration of $\Delta\rho g/\rho$, where $g$ is the local gravity.
They will tend to drift upward
as fast as allowed by their coupling to the surrounding turbulent medium.
The turbulent drag per unit length on a long flux tube moving
upward with a systematic velocity $V_b\hat z$ is
\begin{equation}
F_{drag}=C_d|\vec V_T-\vec V_b|r_t(\vec V_T-\vec V_b).
\end{equation}
If we assume that $|\vec V_b|\ll|\vec V_T|$
then this becomes
\begin{equation}
F_{drag}\approx -C_d{4\over 3} V_T V_b r_t.
\end{equation}
Equating this to the buoyant force we find that
\begin{equation}
V_b \approx {r_tg\over V_T} {3\pi\over 4C_d} {\Delta\rho\over\rho}.
\end{equation}
In the ideal fluid limit $\Delta\rho=\rho$.  Using eq. (\ref{eq:radius})
we get
\begin{equation}
V_b \approx {1\over k_T l_p} {3\pi\over 16} V_T
\left({4E_B\over wE_T}\right)^{1+1/n},
\label{eq:vb}
\end{equation}
where $l_p\equiv P/\rho g$ is the pressure scale height.
In the resistive limit we see from eqs. (\ref{eq:rdiff})
and (\ref{eq:diffrho}) that the product of $\Delta\rho$
and $r_t$ is not a function of the resistivity.
This is a consequence of the
condition for marginally stiff flux tubes.  As a result
the speed with which flux tubes rise is insensitive to
whether or not they are in the resistive regime.

The conventional picture of the magnetic field distribution
is that it tends to segregate from the surrounding gas to
the extent that $V_A$ rises to $V_T$.  For a star the
usual assumption is that flux is lost at a speed of
approximately $V_A\sim V_T$, subject to uncertainties
about removing mass from the magnetic field lines.
We see that this roughly agrees with eq. (\ref{eq:vb})
when the magnetic field is in equipartition with the
turbulent energy density and $k_Tl_p$, which is what we
expect for turbulence driven by a convective instability.
On the other hand, eq. (\ref{eq:vb}) indicates that the
flux tubes that comprise weak magnetic fields will rise
very slowly, with a speed proportional to $V_A^5$ (assuming
that $n=2/3$).  This still leaves open the possibility that
the rate at which magnetic flux rises is dominated by
collective modes or by diffusion.

We will see that a similar degree of agreement between the
predictions of this model and expectations based on a
diffuse field obtains for buoyancy speeds in accretion disks.
This concordance, when $E_B\approx E_T$,
between buoyancy estimates based on
the flux tube model proposed here and the rates derived from
a diffuse field hides some important conceptual differences
between the two pictures.  Vainshtein \& Rosner (1991)
have shown that the conventional picture of magnetic
field distribution leads to the expectation that magnetic
flux is rarely lost from astrophysical objects.  The
model proposed here allows magnetic flux to be lost at the
rate given by dividing $V_b$ by the scale of the system.
Depending on the resistivity of the surrounding gas, mass
is continuously unloaded from the individual flux tubes
and the escape of almost completely empty flux tubes,
bearing significant amounts of magnetic flux, poses no
particular problem.

\subsection{Magnetic Buoyancy in the Sun and Other Stars}

If the turbulence is driven by convection, as we expect
in stars, and the magnetic field is strong, as it appears
to be in the Sun, then $V_b$ is
a large fraction of $V_T$ and this result can only be
taken as a rough indication of the value of $V_b$.
The picture that this suggests
is one in which the magnetic field of the star is generated
in some turbulent zone of width $\Delta Z$ such that
$\Delta Z\Gamma>V_T$, where $\Gamma$ is the dynamo growth rate.
In this case the magnetic field will grow to equipartition
in this layer.
The flux generated in this zone rises through the convective
layer, gradually breaking up into separate flux tubes and
acquiring structure on smaller and smaller scales as the
scale of the local turbulence shrinks.  The tendency
of the flux loops to acquire more and more small scale
structure should keep the magnetic energy density close
to the kinetic energy density as the flux tubes rise.
A full calculation of how this would work must be dynamical,
in the sense that the speed with which the magnetic flux
tubes is rise is fast enough that we should expect some
deviation from a results derived under the assumption that
the magnetic field is in equilibrium with the local turbulence.
In this paper we will limit ourselves to pointing out some
some of the qualitative features we expect to see based on
our model for MHD turbulence.

How would this work in the Sun?  In fig. (1) we show the
value of $\Psi^{1/2}C_d^{-1}$ as a function of the local
temperature for a mixing
length model of the solar convection zone (\cite{s89}).
We note from eq. (\ref{eq:diffrho}) that when this is less
than one it is the fractional
density depletion (times $C_d$)
within a flux tube when $\Psi<1$ (assuming $E_B\approx E_T$).
When it exceeds one it is roughly 2 divided by the fraction of the
flux tube volume occupied by the skin layer.  In the spirit of
mixing length theory we have assumed that the local pressure
scale height is the diameter of a turbulent eddy (or half
the dominant wavelength).  We have also calculated the
resistivity as though the solar plasma were entirely
ionized, which is only a crude approximation near the surface.
We see that for the Sun $\Psi^{1/2}C_d^{-1}$
runs from a few times $10^{-3}$ at the base of the
convection zone to greater than $1$ near the solar surface.
However, the tubes are significantly evacuated only near the
surface, at temperatures less than $\sim 18,000$ K.  For
small values of $C_d$, say $0.1$, the temperature at which the tubes become
evacuated drops below $10^4$ K.  We note that there is evidence that
the SuperFine Structure (SFS) of the Sun consists of unresolved
flux tubes whose magnetic pressure is roughly equal to the ambient
pressure (for a review see \cite{vbt93}, or \cite{s94}).  The
novel feature of this picture is that the existence of largely
evacuated flux tubes on the surface of the Sun would appear to
be a coincidence, marginally achieved on the Sun, and not necessarily
to be expected for stars with significantly different structure.
Of course, none of this should be taken as directly contradicting
the idea that large flux tubes that penetrate the photosphere can
undergo convective collapse (\cite{p78}, \cite{s83}).  It may be
the latter process happens independently of any of the mechanisms
discussed in this paper.

Given that the magnetic field in the bulk of the solar convection
zone is in the resistive regime, the magnetic flux per flux tube
is given by eqs. (\ref{eq:scale}), (\ref{eq:rdiff}) and (\ref{eq:btres})
as
\begin{equation}
\Phi_t=\sqrt{8\pi P} \pi\left({\eta\over k_T V_T}\right)\Psi^{1/4}
\left({4E_B\over wE_T}\right)^{{5\over 4n}+{1\over8}}.
\label{eq:sflux}
\end{equation}
In the narrow layer where the magnetic field is in the
ideal fluid regime this becomes (from eqs. (\ref{eq:bideal})
and (\ref{eq:radius}))
\begin{equation}
\Phi_t=\sqrt{8\pi P} \pi \left({C_d\rho V_T^2\over 4Pk_T}\right)^2
\left({4E_B\over wE_T}\right)^{2+{2\over n}}.
\label{eq:sflux1}
\end{equation}
$\Phi_t$ is shown for the solar convection zone
as a function of the local temperature in fig. (2)
assuming equipartition.
We note that the flux per tube drops monotonically from the base
of the convection zone to point where the resistive regime ends,
and rises thereafter.  If the properties
of the local flux tubes stay in equilibrium with the surrounding
turbulence then each flux tube at the base of the solar convection
zone will make $\sim 10^2$ flux tubes near the layer where
$\Psi\sim1$, and $\sim 20$ flux tubes in the photosphere.
In the same spirit we expect from eq. (\ref{eq:radius}) that
the flux tubes making up the SFS have radii of about 1.3 km.
More realistically we should expect our assumption of strict
equilibrium with the local turbulence to break down near the
top of the convection zone and interpret this as a prediction
of flux tube radii on the order of a kilometer or so in the top
of the convection zone.  As the magnetic flux tubes rise through
the comparatively thin layer separating the top of the convection
zone from the photosphere they will tend to aggregate (since larger
flux tubes will be more buoyant) and speed up, giving rise to
an exponentially decreasing mean magnetic energy density and a
slightly greater typical flux tube radius.

Is our assumption of equipartition justified?
Let's consider the ratio of magnetic energy density to turbulent energy
density as a function of height.  As the magnetic field lines rise
their total flux is conserved.  However the ratio of size of the
local turbulent cells to the original scale of organization of
the field drops sharply.  If we assume that the field is disordered
on intermediate scales, as a consequence of bending on those scales
as the flux tubes rise, then the flux threading
a turbulent cell is proportional to the length of the cell
divided by the local buoyant velocity.  In other
words
\begin{equation}
\Phi_{tot}\sim \left({l_p r_0\over r l_0}\right)\Phi_0
\left({rl_0\over r_0 V_bt_0}\right)propto {l_p\over V_b},
\label{eq:q2}
\end{equation}
where the subscript `$0$' denotes the conditions at the base of the
convection zone, where the dynamo operates, $t_0$ is the characteristic
time for flux tubes to escape from the dynamo region, and the repeated
factor of $r/r_0$ reflects the transverse spreading of flux tubes
imposed by the spherical geometry.  Using eqs. (\ref{eq:fluxa})
and (\ref{eq:vb}) we see from this that in the resistive parts of
the solar convection zone
\begin{equation}
\left({4E_B\over wE_T}\right)\approx \left({\Psi^{1/2}P^{1/2}\over
l_p\rho V_T^3}\right)^{{4n\over 3(2+n)}}.
\label{eq:en1}
\end{equation}
Near the top of the convection zone, where $\Psi\gtrsim 1$
we use eq. (\ref{eq:flux}) instead to obtain
\begin{equation}
\left({4E_B\over wE_T}\right)\approx \left({P^{1/2}\over
l_p\rho V_T^3}\right)^{{n\over 2(1+n)}}.
\label{eq:en2}
\end{equation}
Evaluating eqs. (\ref{eq:en1}) and (\ref{eq:en2}) for the solar
model given in Stix (1989) gives a ratio of magnetic to
turbulent energy that rises monotonically, by a factor of
about 6 between the bottom of the convection zone and the
end of the resistive regime, and by an additional factor
of 30 at the top of the convection zone.  Clearly, if the
dynamo produces a magnetic field in anything like equipartition
with the local turbulence at the bottom of the convection zone,
then the magnetic field will be in equipartition with the
turbulence throughout the convective region, and with less
energy on scales between $l_p$ and $l_0$ than one would expect
from a randomly twisted field on those scales.  This prediction
of approximate equipartition is supported by helioseismological
data (\cite{gmwk91}).

If the large scale poloidal field of the Sun is due to the coherent
generation of magnetic flux in a solar dynamo, which seems
probable given its quasi-periodic oscillations, then the strength
of the field should be related to the magnetic flux which passes
through the dynamo region.  The appropriate measure of the
strength of this field should be the average field strength,
$\langle \vec B\rangle$, in quiet regions
of the Sun, far from eruptions of flux ropes and at an altitude
where $\beta$ is small enough that the magnetic field fills a
large fraction of space.  Choosing high latitudes for comparison
also simplifies the geometry since we need only consider the
flux threading the layer of turbulent cells in the dynamo
region and ignore questions of radial transport of poloidal flux.
Measurements of the magnetic field
strength above the polar regions of the Sun suggest a value
in the range 1 to 2 gauss (\cite{a73}).  If the dynamo takes
place in the first pressure scale height above the bottom
of the convection zone, then the flux per unit area can be
obtained by divided the RHS of eq. (\ref{eq:fluxa}) by $L_T^2$.
An estimate of the surface poloidal magnetic field strength
follows if we multiply this number by $(r_{dynamo}/R_{\sun})^2$.
Following this procedure we obtain a value between $0.6$ and $5$
gauss depending on which fiducial radius near the bottom of the
solar convection zone we
use.  Evidently the dynamo is about as efficient in generating
a large scale poloidal field as it can be, at least at high
latitudes, given the strength of convection in the Sun.

A crude estimate of the distribution of magnetic field energy
as a function of scale can be obtained by calculating the
fraction of the total magnetic energy, as a function of
solar radius, contributed by the flux rising from bottom
of the convection zone assuming that those flux tubes
rise at a fixed fraction of $V_T$ and are essentially
unbent.  The remainder of the magnetic energy will be due to
structures on scales intermediate between $l_0$ and $l_p$.
We can combine eqs. (\ref{eq:btres}) and (\ref{eq:q2})
to get this fraction as a function of $r$.  It is
\begin{equation}
F_{B0}\propto {wB_t\Phi_{tot}\over \rho V_T^2 l_p^2}
\propto {w(r)\sqrt{P}\over \rho V_T^3 r}\Psi^{1/4},
\label{eq:q3}
\end{equation}
In the top layer of the solar
convection zone, where the ideal fluid regime applies, this scaling
should be replaced by one that drops the factor of $\Psi^{1/4}$.
The value of this scaling
factor, normalized to one at the base of the convective zone,
is shown in fig. (3).  We note that without some twisting of
the magnetic field lines as they rise, the ratio of magnetic
to turbulent energy in the photosphere would be very small, and rise sharply
with decreasing $r$.  We see that the steady drop in $F_{B0}$
as $r$ increases implies that each level in the solar convection
zone impresses structure on the magnetic field lines as they
rise, and that the energy contained in magnetic field line
curvature on a scale $L$ is related to the energy necessary to
keep the magnetic energy and turbulent energy in equipartition
as the field lines rise through that layer of the Sun in which
$l_p\approx L$.  We can see from fig. (3) that it is necessary to
stretch the magnetic field lines by a factor of slightly more
than $10^2$, or about $5$ e-foldings as they rise in order to maintain
equipartition.  Since the number of pressure scale heights in the
convective zone is of order a few dozen, and since each flux tube
stays in a given eddy for at least an eddy turnover time,
this is not a unreasonable amount of stretching.

Assuming that the rising magnetic field lines is in approximate equipartition
with the local turbulence, we can estimate the upward flux of
entrained matter by multiplying the volume filling factor of
the magnetic field times $4\pi\rho r^2 V_b$ in the resistive regime,
i.e. most of the convective zone.  In the top of the convection
region, where the ideal fluid regime applies, this must be corrected
by a factor of $\sim (2\sqrt{\eta\tau}/r_t)$, since only the `skin'
of each flux tube is carrying matter.  Using eqs. (\ref{eq:psidef})
and (\ref{eq:btres}) and taking $V_b\sim V_T$ we find
that
\begin{equation}
\dot M\approx {8\pi r^2\rho V_T\over C_d\sqrt{V_T/k_T\eta}}
\end{equation}
in the resistive regime.  The same result (to within a factor of
2 or so) can be obtained in the ideal fluid regime using eq.
(\ref{eq:radius}).  This mass flux drops from $\sim 4\times 10^{20}$
gm/sec at the base of the convection zone to $\sim 1.3\times 10^{19}$
at the top.  In other words, the amount of mass entrained on
rising flux tubes drops by a factor of 30 as they cross the
convection zone, assuming that the flux tubes stay in equilibrium
with their environment.  This is still $\sim 10^{-8} M_{\sun}$ per
year, substantially more than the mass flux in the solar wind,
implying that there is considerable mass unloading from the
flux tubes above the convection zone.

One major omission in this model is that we have treated
the velocity field as sufficiently chaotic that the distribution
of flux tubes can be described entirely in terms of flux tube
interactions.  However, there are slowly shifting convective
cells on the solar surface, i.e. solar granules. It follows
that these relatively stable flow patterns will tend to
collect vertical flux tubes wherever the fluid velocities
are largely vertical, a feature which is beyond the scope of
the simple model described here.  Smaller scale features should
still be described in terms of this model.

Finally we note again that these predictions are all based on the
assumption that the magnetic flux tubes are always able to
reach equilibrium with their environment as they rise.
However, for a stellar magnetic field in equipartition the buoyant velocity
is some large fraction of $V_T$.  It follows that large
deviations from local equilibrium are possible.  They
are less likely for accretion disks, since $V_b$ is typically
much less than $V_T$, except (as we shall see) for thick, or
radiation pressure dominated, accretion disks.

\subsection{Magnetic Buoyancy in Accretion Disks}

In an ionized accretion disk the magnetic field drives the turbulence
through the Velikhov-Chandrasekhar shearing instability
(\cite{v59}, \cite{c61}, \cite{bh91},
\cite{hb91}, and \cite{hgb94}) so that $E_T\sim E_B$ (\cite{vd92}).
The transport of angular momentum in accretion disks via this process is known
as the Balbus-Hawley mechanism.  If we assume that the internally
generated field has a large scale azimuthal component
and a small scale random component induced by the turbulence
then we have $k_T\sim \Omega/V_A$, where
$\Omega$ is the local rotational frequency.  Since
for a disk, $l_p\sim H$, where $H$ is the disk thickness,
and $c_s\sim H\Omega$ the flux tubes drift upward with
a systematic velocity given by
\begin{equation}
V_b \approx {V_A^2\over c_s}\approx {V_T^2\over c_s}\sim \alpha c_s,
\label{eq:diskrise}
\end{equation}
where we have used the fact that the dimensionless viscosity $\alpha$
is approximately $V_T^2/c_s^2$.
In other words, the magnetic flux tubes rise
at a speed which is less than the local turbulent velocity
by a factor of the Mach number.  A similar result
can obtained from a qualitative argument based on the
nonlinear interaction between the shearing and buoyant modes of
a diffuse magnetic
field (\cite{vd92}).  Consequently one predicts that
magnetic flux is lost from the disk at a
rate of $V_A^2/(c_s H)\sim\alpha\Omega$.
Note that we have dropped all constants of order unity in this argument.

So far we have assumed that the ambient pressure is supplied
by charged particles.  The modest resistivity of most astrophysical
plasmas then allows us to propose that the magnetic field
pressure can be very large inside the flux tubes, with a
compensating deficit of gas pressure.  However, in radiation pressure
dominated environments the diffusion of photons into flux tubes will
prevent the magnetic field pressure from ever dominating even small volumes
in the plasma.
This implies large and weak flux tubes which, if effectively
evacuated of matter, will be much more buoyant than a diffuse field would
be.  Consequently the magnetic dynamo in a radiation pressure dominate disk
will saturate at a lower level, giving rise to a smaller effective viscosity.
We can make this point more quantitative by observing that in
this situation $B_t^2$ is limited to $8\pi P_{gas}$.  Assuming equipartition,
we can calculate the typical flux tube radius by truncating the fractal
distribution of flux tubes in an ideal fluid at the scale where the mean
magnetic pressure is at this limit.
Eq. (\ref{eq:radius}) becomes
\begin{equation}
r_t\approx {C_d \rho V_T^2\over 4k_T P_{gas}}.
\label{eq:rdisk}
\end{equation}
Neglecting resistivity we still expect that $\Delta\rho=\rho$,
i.e. these flux tubes are virtually empty.  Therefore
\begin{equation}
V_b \approx {P_{tot}\over P_{gas}}{1\over k_T l_p} {3\pi\over 16} V_T,
\label{eq:vb1}
\end{equation}
or for an accretion disk
\begin{equation}
V_b \sim {P_{tot}\over P_{gas}}\alpha c_s.
\label{eq:vb1a}
\end{equation}
Since the magnetic flux lost to buoyancy must be replaced
we can equate the dynamo growth rate to $V_b/H$, implying
\begin{equation}
\alpha\sim \left({\Gamma_{dynamo}\over\Omega}\right)\left({P_{gas}\over
P_{tot}}\right).
\label{eq:alpha}
\end{equation}
If the dynamo is unaffected by the dominance of radiation pressure,
then this implies that the vertically integrated heating rate in a
radiation
pressure dominated disk is
\begin{equation}
Q_+\sim \alpha P_{tot} c_s\sim \left({\Gamma_{dynamo}\over\Omega}\right)
P_{gas}c_s,
\end{equation}
which depends only on the gas pressure.  This result applies only if
the dissipation rate remains dominated by stresses induced by
the Velikhov-Chandrasekhar instability
and if $V_b$ remains less than $V_T$.

Is it reasonable to treat the radiation pressure as uniform across
the flux tubes?  If we are in the ideal fluid limit then the
speed with which photons will diffuse into a flux tube is
approximately
\begin{equation}
{c\over \sigma_Tn_e\sqrt{\eta\tau}}.
\end{equation}
Since the photons, like the gas, are eliminated from the flux tubes
through the process of stretching, folding, and pinching off loops
it follows that
\begin{equation}
\Delta P_\gamma {c\over \sigma_Tn_e\sqrt{\eta\tau}}r_t^{-1}\approx
{P_\gamma-\Delta P_\gamma\over\tau},
\label{eq:raddiff}
\end{equation}
where $\Delta P_\gamma$ is the photon pressure differential and
$P_\gamma$ is the external photon pressure. In the limit where
$\Delta P_{\gamma}<P_{gas}$ we have $\Delta P_{\gamma}\ll P_{\gamma}$.
Using eq. (\ref{eq:rdisk}) we can rewrite this as
\begin{equation}
\Delta P_{\gamma}\sim {P_{\gamma}\sigma_T n_e\over c k_T}
\sqrt{\eta\Omega}{C_d\over 4}\left({\rho V_T^2\over P_{gas}}\right)
\end{equation}
Since $V_T^2\approx\alpha c_s^2$ this is
\begin{equation}
\Delta P_{\gamma}\sim Re_B^{-1/2}\alpha^2 {P_\gamma^2\over P_{gas}}
{\sigma_T n_e Hc_s\over c},
\end{equation}
where $Re_B$ is the magnetic Reynolds number and
we have discarded $C_d/4$ as a
factor of order unity.  For radiation pressure dominated disks
\begin{equation}
\sigma_Tn_eH \sim{c\over c_s\alpha},
\end{equation}
which implies that
\begin{equation}
\Delta P_\gamma\sim Re_B^{-1/2} \alpha P_{gas}\left({P_\gamma\over P_{gas}}
\right)^2.
\end{equation}
Finally, in assuming that we were in the ideal fluid limit, as we
did at the beginning of this discussion, we implied that
$Re_B\alpha^2\gg P_{gas}^2/P_\gamma^2$.
This in turn implies that
\begin{equation}
\Delta P_{\gamma}\ll P_{gas}\alpha^2\left({P_\gamma\over P_{gas}}\right)^3
\sim P_{gas}\left({\Gamma\over\Omega}\right)^2 {P_\gamma\over P_{gas}}
\sim {\Gamma\over\Omega}\left({V_b\over V_T}\right)^2.
\label{eq:raddiff2}
\end{equation}
We conclude that in the ideal gas limit $\Delta P_\gamma\ll P_{gas}$
unless $V_b\gg V_T$. However, in this limit our assumption regarding
the ordering of the velocities is violated.  Moreover it is unclear
that one can apply the Velikhov-Chandrasekhar instability to this
case.  Fortunately, this limit only obtains when $P_\gamma/P_{gas}$
exceeds $\Omega/\Gamma_{dynamo}$.  In the internal wave driven dynamo
model (\cite{vjd90}, \cite{vd92}, and \cite{vd94})
the angular momentum deposited by the nonlinear dissipation
of the internal waves gives a minimum for $\alpha$ which will dominate
over the value derived from the Velikhov-Chandrasekhar instability
at smaller values of $P_\gamma/P_{gas}$.

In the resistive limit eq. (\ref{eq:raddiff}) becomes
\begin{equation}
\Delta P_\gamma {c\over \sigma_Tn_e\eta\tau}\approx
{P_\gamma-\Delta P_\gamma\over\tau}.
\label{eq:raddiff3}
\end{equation}
Following the same line of reasoning as above we can replace
eq. (\ref{eq:raddiff2}) with
\begin{equation}
\Delta P_\gamma\sim {P_\gamma\over Re_B}\sim P_{gas}
{\eta\Omega\over\alpha c_{gas}^2},
\end{equation}
where $c_{gas}^2$ is the sound
speed of the hot gas alone (i.e. $P_{gas}/\rho$). Assuming that the
magnetic diffusivity is dominated by electron-ion collisions then
$\eta/c_{gas}^2\approx 2\times 10^{-11} T_6^{-5/2}$ seconds.  Consequently,
for AGN, for which $T_6$ is of order unity, $\alpha$ is of order $10^{-2}$
to $10^{-3}$, and $\Omega$ is no more than $10^{-3}$, we conclude that
$\Delta P_\gamma\ll P_{gas}$ even if the resistive limit applies.
Disks around smaller mass black holes will tend to have lower disk
temperatures (by roughly a factor of $M_{BH}^{1/4}$) and larger
values of $\Omega$ (by a factor of $M_{BH}^{-1}$) so even if we
consider a solar mass black hole we still find that $\Delta P_\gamma$
will be less than $P_{gas}$ by at least a factor of $10^{-8}$.
We conclude that our assumption that the radiation pressure does
not vary significantly across a flux tube boundary is satisfied for
all accretion disks likely to be dominated by radiation pressure.

In fact, we can show that it is quite likely that such disks are always
in the ideal fluid regime.  Taking into account that the magnetic pressure
is limited to the gas pressure alone in such disks, eq. (\ref{eq:diffcrit})
can be modified to yield a criterion
for the resistive regime.  It is
\begin{equation}
\left({\alpha P_\gamma\over P_{gas}}\right)^3
\left({c_{gas}^2\over \Omega\eta}\right)<\sim16.
\label{eq:diffcrit1}
\end{equation}
Taking into account eq. (\ref{eq:alpha}) and assuming that the
resistivity is dominated by electron-ion collisions this implies
that
\begin{equation}
\left({\Gamma_{dynamo}\over \Omega}\right)^3 \Omega^{-1} T_6^{5/2}
<\sim 3\times10^{-10},
\label{eq:diffcrit2}
\end{equation}
where $\Omega$ is given in radians per second.  Since
$\Gamma_{dynamo}/\Omega\sim\alpha$ for normal, i.e. gas pressure
dominated, accretion disks, and
phenomenological determinations of $\alpha$ in such disks tend
to give
values in the range $10^{-1}$ to $10^{-2}$  this makes it
seem relatively unlikely that a realistic model of a radiation
pressure dominated accretion disk could be in the resistive
regime.

We still need to determine whether or not the magnetic dynamo is
suppressed by viscosity in radiation pressure dominated disks.  Applying
eq. (\ref{eq:diffvis}) to accretion disks and once again remembering
that the matter pressure, rather than the total pressure, limits
the local magnetic field strength, we find that the criterion for
ignoring viscosity is
\begin{equation}
4\pi^3< \left({\alpha P_\gamma\over P_{gas}}\right)^2 {c_{gas}^2\over
\Omega\nu},
\label{eq:z9}
\end{equation}
where we have ignored $C_d$ and $\gamma$ as begin close to unity.
The viscosity is apt to be dominated by the photon shear viscosity
which is (\cite{t30}, \cite{smt71})
\begin{equation}
\nu={8\over 9} {P_\gamma\over\rho c^2} {c\over n_e\sigma_T}.
\end{equation}
Once again discarding constants of order unity we see that
since
\begin{equation}
{P_\gamma c\over n_e\sigma_T H}\sim\alpha\Sigma c_s^2\Omega,
\end{equation}
where $\Sigma$ is the matter column density in the disk, it follows that
\begin{equation}
\nu\sim {\alpha H c_s^3\over c^2}.
\label{eq:z10}
\end{equation}
Combining eqs. (\ref{eq:z9}) and (\ref{eq:z10}) we find that viscosity
will not dominate the magnetic field dynamics when
\begin{equation}
4\pi^3< \sim \left({\alpha P_\gamma\over P_{gas}}\right)\left({c\over
c_s}\right)^2,
\label{eq:z11}
\end{equation}
or
\begin{equation}
4\pi^3< \sim {\Gamma_{dynamo}\over \Omega}\left({c\over r\Omega}\right)^2
\left({r\over H}\right)^2.
\label{eq:z12}
\end{equation}
If the magnetic dynamo is an $\alpha-\Omega $ dynamo
driven by purely local processes, then
$\Gamma_{dynamo}/\Omega$ is some number less than one.
In the internal wave driven dynamo model $\Gamma_{dynamo}\sim (H/r)^k\Omega$,
where $k$ is between 1 and $1.5$, a result which is also consistent
with
a number of phenomenological studies of accretion disk models
(\cite{mo83}, \cite{mm84}, \cite{m84}, \cite{s84}, \cite{mo85},
\cite{lpf85}, \cite{cwp86}, \cite{hw89}, \cite{mw89}, \cite{mwa89},
and \cite{c94}).
In either case we see that whereas the right hand side of eq. (\ref{eq:z12})
is apt to be quite large in most disks, near the event horizon of
a black hole accreting near the Eddington limit, so that $r\Omega$
can approach $c$ and $H$ can approach $r$, this inequality may
not be satisfied.  In other words, near the very inner edge of
accretion disks around black holes the magnetic dynamo could fail
altogether due to photon viscosity.

One consequence of these results is that we can estimate the
fraction of the energy generated by dissipation within the
disk which is ejected in the form of rising magnetic flux tubes.
The magnetic energy density, $E_B$, is roughly $\alpha P_{tot}$.
The magnetic energy flux is just
\begin{equation}
F_B\sim\ \alpha P_{tot}V_b\sim \alpha P_{tot}c_s\left({V_b\over c_s}\right),
\label{eq:z6}
\end{equation}
where $\alpha P_{tot} c_s$ is the vertically integrated energy
generation rate (and therefore approximately equal to the radiative
flux from the disk).  Comparing eqs. (\ref{eq:diskrise}) and eq. (\ref{eq:z6})
we see that the fraction of energy carried away by rising flux tubes
is roughly $\alpha$ for a normal disk.  This energy is likely to
be eventually dissipated as nonthermal radiation from the disk
chromosphere and corona.  For a radiation pressure dominated disk
eq. (\ref{eq:diskrise}) must be replaced by eq. (\ref{eq:vb1a}) and
the fraction of a disk's energy budget carried by rising flux tubes
is $\alpha P_\gamma/P_{gas}$ or $\Gamma_{dynamo}/\Omega$.  The latter
expression
will be approximately correct regardless of whether or not the
disk is dominated by radiation pressure.

We conclude that unless $P_\gamma>P_{gas}(\Gamma_{dynamo}/\Omega)^{-1}$
we can model a radiation pressure dominated disk using a dimensionless
viscosity which couples only to the gas pressure, provided that
purely hydrodynamic effects do not contribute a significant
viscosity (as they will in the internal wave driven dynamo model).  Since
$\alpha\sim\Gamma_{dynamo}/\Omega$ in a gas pressure dominated disk this
is equivalent to limiting the radiation pressure to a value no more
than two or three orders of magnitude greater than the gas pressure.
However, this is less of a limit than it might appear, since coupling
dissipation to gas pressure alone causes the ratio $P_{\gamma}/P_{gas}$
to rise quite slowly with decreasing radius.

In a sense this result is anti-climactic.  This kind of
model has been previously proposed (\cite{le74})
as a way to avoid the severe instabilities which would otherwise
occur in an accretion disk with a local viscosity coupled to the
total pressure (\cite{prp73}, \cite{le74}, \cite{ss76}).
In fact, magnetic buoyancy has been specifically cited as a mechanism
which might limit dissipation to a rate proportional to the
gas pressure rather than the total pressure (\cite{el75}, \cite{c81},
\cite{sc81}, and \cite{sr84}).
In a similar vein, Sakimoto \& Coroniti (1989) claimed that any model
for angular momentum transport due to global magnetic stresses
proportional to the total pressure could not be internally self-consistent.
However, their model assumed that angular momentum transport
was due to global Reynolds stresses rather than the
Velikhov-Chandrasekhar instability.  Moreover,
they lacked any clear criterion for the flux tube radius.
Consequently, their result was expressed as a preference for
coupling to gas pressure rather than total pressure, given
a choice between the two, rather than a derivation of the
correct coupling.  What we have shown here is that given our model
for MHD turbulence, and the assumption that the Balbus-Hawley mechanism
is responsible for angular momentum transport in accretion disks,
the dissipation rate is proportional to the product of the
dynamo growth rate and the gas pressure.

\section{Conclusions}

We have proposed a model for the distribution of the magnetic field in
a highly conducting, turbulent medium with a high $\beta$.  The basic
feature of the model is that the magnetic flux is distributed in
bundles of small radius and large Alfv\'en velocity.  The typical
scale of flux tube curvature is the scale at which the turbulent
kinetic energy density and the average magnetic field energy density
are in equipartition.  This is much larger than the typical flux tube
radius, which is set by the condition that the tubes be marginally
stiff to fluid motions on the curvature scale. The skin depth of the
flux tubes can be smaller still, depending on the regime begin
considered.  The direction of the magnetic field inside the flux tubes
is strongly correlated over all scales less than the curvature
scale, i.e. the magnetic field does not show numerous reversals
on small scales.  This structure
implies efficient reconnection, allows the magnetic field and
the bulk of the plasma to move independently, and yet retains
enough coupling between the two that the basic notion of a
fast dynamo remains plausible.  We have not attempted to
rederive mean-field dynamo theory from this model, nor find
a replacement for it based on the dynamics of the flux tubes.
We note only that the basic features of mean field dynamo theory,
twisting due to forcing by the surrounding fluid flow and
reconnection, are inevitable parts of this model.
Applying the model to stars we
can see that the kind of substructure observed in the sun is
the inevitable result of a dynamo buried at the base of the
convective zone.  Applying the model to accretion disks we see
that, as previously claimed, magnetic flux loss from accretion
disks is relatively inefficient and proceeds at a rate that
scales with the rate for vertical turbulent diffusion.  In addition,
we have found that for values of $P_\gamma/P_{gas}$ moderately
greater than one the dissipation couples only to the gas pressure.

We have made no attempt to apply this model to the galactic magnetic
field.  There are several reasons for this.  First, the mean
magnetic pressure in the disk of the galaxy probably exceeds the thermal
pressure, although it is in rough equipartition with the turbulent
pressure and the cosmic ray pressure.  This violates our assumption
of large $\beta$.  Second, the disk is filled with supersonic turbulence
whereas we have taken $V_T/c_s$ as a small parameter.  Third, the
magnetic field of the galaxy interacts with both the gas in the disk,
and the cosmic rays.  It is not clear to what extent the latter can
be treated as a fluid, nor what their role might be in the galactic
dynamo.  Fourth, the galactic disk is a highly inhomogeneous environment,
with many local sources of outflow, complete with entrained particles and
fields.  This makes it unlikely that the model we have presented here,
based on turbulent cells whose internal properties are statistically
homogeneous, can be applied.

It is intriguing to note that the proposed distribution of magnetic
flux in an ideal, turbulent fluid seems to maximize the
dissipation of the
turbulent energy, at least in the ideal fluid regime.  A qualitative
argument along these lines is as follows.  Consider a turbulent cell
threaded by some fixed amount
of flux.  The rate at which turbulent energy is dissipated in an
unmagnetized eddy is fixed by the large scale eddy turnover rate.
In order to enhance this rate of dissipation the magnetic field
needs to absorb energy directly from the large scale eddies and
transfer it to some much smaller scale on a short time scale.
There are basically two ways this could happen.  First, the magnetic
field might be gathered into flux tubes which are stiff, on
the largest scale of turbulence.  In this case kinetic energy
in the large scale eddies is dissipated in the turbulent wakes
behind the flux tubes on a time scale $\sim V_T/r_t\gg V_T/L_T$.
Second, the magnetic field might be
pliant on large scales, but constantly reconnecting on smaller
scales so that energy absorbed by the field is immediately
transferred to smaller scale field loops which rapidly collapse
or are folded into smaller and smaller loops.

In the first
case the energy dissipation rate induced by the magnetic field
is
\begin{equation}
\dot E_T\approx N_T \rho V_T^2 (r_t^2L_T){V_T\over r_t},
\end{equation}
where $N_T$ is the total number of flux tubes.
The constraint imposed by the external magnetic flux can be
expressed in this case as $N_TV_{At} r_t^2\equiv \Phi_T$, where
$\Phi_T$ is a constant.
Consequently,
\begin{equation}
\dot E_T\approx {\rho V_T^3 L_T\Phi_T\over r_tV_{At}}.
\label{eq:z1}
\end{equation}
The condition that these flux tubes be stiff is, ignoring
constants of order unity, the condition that
\begin{equation}
V_{At}^2r_t\ge V_T^2L_T.
\label{eq:z2}
\end{equation}
Comparing eqs. (\ref{eq:z1}) and (\ref{eq:z2}) we see that the
energy dissipation rate is maximized if (\ref{eq:z2}) is just
marginally satisfied and if $V_{At}$ is as large as possible,
i.e. $V_{At}\approx c_s$.  In this limit the energy dissipation
rate due to the presence of the magnetic field is
\begin{equation}
\dot E_T\approx \rho c_s V_T L_T\Phi_T.
\label{eq:z3}
\end{equation}

In the second case the magnetic field will absorb, and dissipate
energy, at a rate proportional to its total energy times the shear
due to large scale flows, i.e.
\begin{equation}
\dot E_T\approx E_B {V_T\over L_T}.
\end{equation}
The problem of maximizing the energy dissipation rate is equivalent
to maximizing the magnetic energy density subject to the constraints
implied by $\Phi_T$ and the dynamics of the turbulence.  We can
express the total magnetic energy as
\begin{equation}
E_B\approx \Phi_l l V_{At}\left({L_T\over l}\right)^3,
\end{equation}
where $l$ is the scale on which the flux tubes can resist
the local fluid motions, and $\Phi_l$ is the magnetic flux,
divided by a factor of $(4\pi\rho)^{1/2}$, across a typical
volume of size $l$.  Since the magnetic field lines are essentially
random walks on larger scales we can use eq. (\ref{eq:phran})
to obtain
\begin{equation}
\dot E_T\approx \Phi_T V_{At}V_T\left({L_T\over l}\right).
\end{equation}
Comparing this result to eq. (\ref{eq:z3}) we see that if
$V_{At}=c_s$ then weaker and more numerous flux tubes dissipate
energy more efficiently than a few rigid flux tubes, provided
the more numerous flux tubes do not interfere with the turbulent
cascade.  The latter condition is particularly important, since
a uniform magnetic field with an energy density greater than
the energy density of local eddies can be shown to strongly
inhibit energy dissipation (\cite {k65}, \cite{dc90}) by replacing
the usual turbulent cascade with Alfv\'enic turbulence.  When
the number of flux tubes in turbulent eddies of size $l$, $N_l$
exceeds $l/r_t$, then even if the flux tubes are not completely stiff on
this scale, their mutual shadowing implies that most of the energy
on this scale goes into the magnetic field, which can dissipate
this energy only through Alfv\'enic turbulence.  Mutual shadowing
is moderate if $N_lr_t\lesssim l$.  Since
\begin{equation}
\Phi_l\sim N_l V_{At} r_t^2\sim V_{At}r_t l\sim \Phi_T {l\over L_T},
\end{equation}
it follows that
\begin{equation}
V_{At}r_t \sim {\Phi_T\over L_T}\approx {V_l^2 l\over V_{At}},
\end{equation}
where the last expression is just a restatement of the fact that
$l$ is defined as the scale of marginal stiffness.  The ratio
of $V_{At}$ to $l$, which must be maximized to maximize energy
dissipation, is therefore proportional to $V_l^2$.  We conclude
that the energy dissipation rate is maximized when $l$ is maximized.
Our last equation implies that $V_l^2l$, and therefore $V_l^2$,
is maximized when $V_{At}$ is maximized, which brings us back to
the condition that $V_{At}\approx c_s$.  Combining this with
our previous results we find
\begin{equation}
E_B\sim \rho V_l^2 L_T^3,
\end{equation}
\begin{equation}
\Phi_T\sim {lV_l^2 L_T\over c_s},
\end{equation}
and
\begin{equation}
r_t\sim {\Phi_T\over L_Tc_s}\sim l{\cal M}_l^2.
\end{equation}
Adding the assumption that $V_l^2\propto l^n$ will allow us to rederive,
in less exact form, the major results of \S II.  The only remaining
result is the correlation between flux tubes on scales smaller than
$l$.  This plausibly follows from the condition that the turbulent
wakes of the individual flux tubes can be made maximally efficient
at dissipation if they overlap at every scale.

It appears that the model proposed here is equivalent to claiming
that the magnetic field in a turbulent flow is an example of a
{\it dissipative structure}, an ordered state which promotes the
dissipation of energy and the production of entropy.
This presents a sharp contrast to the
effect of a more smoothly distributed magnetic field (\cite {k65}, \cite{dc90})
which inhibits the decay of turbulent eddies and their
eventual dissipation.

This model contains several implications for numerical MHD calculations.
First, a simulation which starts from a diffuse magnetic field (e.g.
\cite{hgb94})
embedded in a turbulent fluid will show a strong initial growth
in the mean magnetic energy density, {\it even if there is no
dynamo at work} with an e-folding rate close to the eddy turnover rate.
If the calculation is compressible with a
Mach number close to one (in the sense of satisfying eq. (\ref{eq:diffcrit}))
then the final magnetic energy density will be roughly the geometric
mean between its initial value and the thermal pressure of the
surrounding fluid times $w$, i.e. times a factor of order three or four.
A true dynamo can be distinguished from this effect only by a
careful examination of the large scale distribution of magnetic
flux, or by starting from a state consisting of flux tubes
with $V_A\sim c_s$.  Claims regarding the ability of the Velikhov-Chandrasekhar
instability to support a dynamo should be evaluated with this point
in mind, especially since current simulations show a strong dependence
on the initial state, suggesting that no dynamo is present.
However, given the current state of three dimensional
MHD codes eq. (\ref{eq:diffvis}) represents a major obstacle to producing
realistic simulations.
The minimum value of ${\cal M}^2$ which avoids the viscous regime
is of order $120$ divided by the Reynolds number.  This means
that the current generation of codes are limited to very slightly
subsonic turbulence, or a value of the resistivity high enough
to satisfy eq. (\ref{eq:svmr}).  We have already noted that
satisfying this criterion is also difficult.  The presence of
a large scale shear evidently prevents the strongly viscous
regime from producing a completely stagnant situation (\cite{hgb94}),
nevertheless the flux tube dynamics should still be very different in
this regime.  The field amplification due to flux tube formation
is unlikely to play a major role in
astrophysical objects, since the initial state is normally unrealistic.
(The early galaxy may be one exception to this rule.)
Second, the failure to create a simulation in which the fluid is in the
resistive or ideal fluid regimes will prevent the magnetic energy
density from reaching equipartition with the turbulent flow, even
in the presence of a strong dynamo.  Current MHD simulations of stellar
convection suffer from this difficulty, and should not be taken as
realistic models of stellar dynamos.  Third, such failures can
occur even for large values of the magnetic and fluid Reynolds
numbers.  Success is most likely for compressible codes with large
values of $V_T/c_s$, i.e. not too much less than one, or for
incompressible codes with $\eta/\nu\gg 1$.  In particular, one can
go from the viscous regime to the resistive regime for a simulation
of incompressible MHD turbulence most easily by allowing the value
of $\eta$ is be much larger than the minimum value given by the
nature of the computer code.  Improving the code by lowering
$\eta$ can actually result in {\it less} realistic results.

On the other hand, current numerical codes are likely to prove
essential in testing the model proposed in this paper, even if they
can't be used to simulate realistic situations.  The model proposed
here makes some specific predictions concerning the saturated state
of the magnetic field in simulations with strong dynamos.
(The presence of a dynamo may be necessary to prevent the magnetic
field from disappearing when the imposed magnetic flux is very small,
or zero).  In particular, we predict
the existence of a critical Reynolds number $\sim 60/C_d$ (where the
Reynolds number is defined with the inverse wavenumber of the strongest
flows as the fiducial length).
Simulations with smaller Reynolds numbers should show flux tubes
with internal Alfv\'en velocities $\sim V_T$, and a flux tube radius
(or alternatively, the magnetic Taylor
microscale) between $C_d/2k_T$ and $30 \nu/V_T$.  Simulations that
start with a very weak uniform magnetic flux will evolve toward
a state in which each large scale eddy has one flux tube, with the
minimum flux tube radius, and an average magnetic field energy which
is a fraction, approximately equal to $0.04C_d^2$, of the kinetic
energy density.  If the imposed magnetic flux is too large
to be accommodated within so few flux tubes, then the magnetic energy
density will exceed its minimal value by the ratio of the total
magnetic flux to the maximum value consistent with the minimal state.
Initially, the increased magnetic energy will be reflected in a
proportional increase in the area of each flux tube (or an increase
in the magnetic Taylor microscale proportional to the square root
of the flux).  However, once the average flux tube radius reaches
its maximum value it will stabilize and any further increases in
magnetic energy will be accommodated in additional flux tubes of
the same large size.  We note again that the ratio of the maximum
and minimum flux tube radii is inversely proportional to the
Reynolds number of the simulation.  The value of the resistivity
will have little effect on these predictions as long as the magnetic
Reynolds number is large, and the resistivity is insufficient to
damp the dynamo before the field has had a chance to form flux tubes.
As we move to higher Reynolds numbers, i.e. larger than the critical
value cited above, we should see both the ratio of the average
magnetic energy density to the turbulent kinetic energy density,
and the ratio of the magnetic Taylor scale to the eddy size, fall off inversely
with the Reynolds number.  The maximum magnetic flux consistent
with $\sim 1$ flux tube per eddy will drop
at the slightly faster rate of $Re^{-3/2}$ as the Reynolds number is
increased.  Initial states with greater magnetic flux will saturate
at a higher magnetic energy density (scaling linearly with the initial magnetic
flux) but with flux tubes of the same radius.
Everywhere in the strongly viscous regime, which is the one accessible to
current numerical simulations, the radius of curvature of the field lines
will be $\sim L_T/4$.  The contrast between this scale, which may be
approximated by
$[\int B^4dV/\int ((\vec B\cdot\vec\nabla)\vec B)^2 dV]^{1/2}$,
and the magnetic Taylor microscale will become increasingly obvious
for Reynolds numbers above the critical value.
The simulations of Nordlund et al.
(1992) and Tao et al. (1993) are consistent with these predictions
(modulo some uncertainty about the actual effects of varying of
$\eta$) but have similar Reynolds numbers, bracketing the dividing
line given in eq. (\ref{eq:modvisc}).

The division between large and small scale magnetic fields, an
initial step in traditional mean-field theory, is not particularly
useful here.  The
small scale features of this model, the flux tube radius and the
flux tube skin depth, are intimately connected to the
dynamics of the large scale field.  This makes it difficult to
compare this treatment of magnetic field dynamics to recent work
in mean-field theory (e.g. \cite{cv91}, \cite{gd94}).  However, the two
approaches
do have different predictions for numerical simulations. For example,
Gruzinov \& Diamond predict that the dynamical behavior of the
large scale magnetic field will change dramatically once its amplitude
exceeds $(\rho V_T^2/R_m)^{1/2}$, while the small scale magnetic field
reaches equipartition with the turbulent energy density.  It does
not predict the existence of a critical value of the Reynolds number,
or a decline in $E_B/E_T$ at high Reynolds number.  Moreover, it
suggests a final state which is quite sensitive to $\eta$.  It
follows that a confirmation of the predictions in the preceding
paragraph would be strong argument for using the flux tube
dynamics suggested here, rather than mean-field theory and the
failure of those predictions, and a confirmation of the predictions
of Gruzinov and Diamond, would have the opposite effect.

It is appropriate to pause at this moment and remember what is not included
here.  We have included only the most basic features of fluid turbulence,
meaning that we have characterized the fluid motion by an large scale
eddy size and the velocity on that scale.  We have argued that the
smaller eddies play very little role in the dynamics of the magnetic
field, except for the turbulent wakes generated by the flux tubes
themselves.  Real turbulence is often characterized by intermittent
coherent structures.  We have made no allowance the effects of such
structures and make no predictions concerning simulations which
explicitly include coherent flows.
Realistic astrophysical situations may include cases
where turbulent motion is driven on a variety of scales.  Our results
suggest that the most important scale is the one where most of the
turbulent energy resides, at least for magnetic fields in equipartition
with the turbulence, but in cases where the energy peak is broadly
distributed, or where there are two or more competing peaks in the Fourier
spectrum, the magnetic field may show more complicated structure.
Nevertheless, within the limitations imposed by our neglect of these
points, our model appears to explain many of the features observed in
numerical situations and some aspects of the solar magnetic field.
It is appropriate to regard it as a crude sketch of a more complete
theory.

\acknowledgements

This work was supported by NASA grant NAGW-2418.  I would like to
thank several people for useful discussions, including Fausto
Cattaneo, Jung-Yeon Cho, Patrick Diamond, Robert Duncan, Russell Kulsrud,
Norman Murray, Stefano Migliuolo,
Christopher Thompson, and Samuel Vainshtein.  I would also like to
thank the anonymous referee of a previous, unpublished note,
who persuaded me that magnetic buoyancy cannot be understood without
a theory for predicting flux tube radii under different physical
conditions.  The initial impetus
for this work came from a visit to the Canadian Institute for
Theoretical Astrophysics.

\appendix

\section{A Comparison to Numerical Work}

In a recent paper Tao et al. (\cite{tcv93}) simulated three dimensional
incompressible MHD turbulence with an imposed helicity.  The simulation took
place in a box with sides of length $2\pi$.  The fluid had $\nu=\eta=1/130$.
The turbulence was driven by imposed bulk forces tuned so that the
rms fluid velocity was one.  The turbulence was supported by forces
distributed in phase space from $|k|=2$ to $|k|=4$.  They found that
the magnetic field energy density tended to saturate at values far
below equipartition with the fluid kinetic energy. Here we show that
their simulation was solidly inside the viscous regime, and that their
results can be understood in terms of the formulae given in this paper.

We begin by noting that $c_s=\infty$ so that ${\cal M}=0$.  Taking
$k_T\approx 3$ we note that
\begin{equation}
Re\equiv\left( {V_T\over k_T\nu}\right)\approx 43.
\end{equation}
Comparing to eq. (\ref{eq:modvisc}) we see that this places us in,
or close to, the regime of moderate Reynolds numbers where we expect
the Alfv\'en velocity in the flux tubes to be close to $V_T$.
Following eq. (\ref{eq:q1}) resistivity will be negligible provided
that the Reynolds number exceeds $4/C_d^2$, which will certainly be
true here.  It follows that we are well inside the viscous regime.

In all of their simulations they started with a uniform magnetic
flux.  In the first two cases the starting value of $V_A$ was
$32^{-1}$ and $256^{-1}$ and the simulations evolved towards the
same final state, which we can identify with our minimal viscous
state.  In the last case they took an initial value of $30^{-1/2}$,
and reached a somewhat different state.

We may anticipate their results to the point of noting that
the typical magnetic Taylor microscale they found in their minimal
state was $\sim 0.2$, which should be close to the typical
flux tube radius.  The appropriate value of $C_d$ for the
flow past the flux tubes is $\sim 1.5$ (\cite{r46}).  This implies
that the dividing line for the moderate Reynolds number regime
is (from eq. (\ref{eq:modvisc})) at $Re=41$.  In other words,
we expect the minimal flux tubes to be close in size to the
maximal flux tubes (given in eq. (\ref{eq:rvisc})) with a typical
size of
\begin{equation}
r_t\approx 0.25
\end{equation}
We note that this is 25\% larger than the observed value, but considering
the crudity of our calculations this counts as excellent agreement.
The minimal state should have one flux tube per turbulent cell, for
a total energy density of
\begin{equation}
E_B\approx {w\over 2} V_T^2 {\pi r_t^2\over L_T^2}\approx 0.044,
\end{equation}
where we have taken $w=2$, the appropriate value for the viscous regime.
The observed value is $E_B=0.05$.  We note
that the flux per turbulent cell in the minimal state (defined
in terms of the area integral of $V_A$) is 0.13.  The sign of this
flux will vary from one turbulent cell to the next.  However,
if the initial value of $V_A$ exceeds $\sim 0.05$ (which is
the lower limit from eq. (\ref{eq:zz}) for this simulation)
then the simulation
will be unable to fit its total flux into a minimal state configuration.
In other words, the simulation with $V_A=32^{-1}$ initially is
below the dividing line by a factor of about 1.7.
The simulation with $V_A$ initially
set to $30^{-1/2}=0.18$ will, according to our predictions, settle
into a state with flux tubes of approximately the same size, but
with an energy density equal to $(w/2) V_T\Phi_{tot}$, or
$E_B\approx 0.18$.  Tao et al. find an energy of about $0.18$ and
a radius of about $0.3$.  The fact that the area per flux tube
goes up in this case by a factor of about 2, whereas the total
magnetic energy rises by a factor of almost 4, is due to the increased
number of flux tubes per turbulent cell.

The discrepancy in radius in this third case
can be understood as following from this simulation being slightly
deeper into the regime of moderate Reynolds numbers than we have
estimated, so that a larger initial flux increases the flux tube
radius slightly.  Since the minimal and maximal flux tube radii differ by
a factor proportional to $\nu$ in this regime, we expect that the
two will converge for a similar simulation with $\nu\approx 0.005$,
or smaller by a factor of $3/2$.  To be more specific, we expect that
the larger of the two radii will shrink down to the smaller as $\nu$
decreases.  Still smaller values of $\nu$ should
produce an Alfv\'en velocity in the flux tubes which is larger than
$V_T$ by a factor that rises inversely as the square root of $\nu$,
while the flux tube radius falls with $\nu$.  If $\nu$ becomes
small enough, then the inequality in eq. (\ref{eq:svmr}) will be satisfied
and the simulations will finally reach the resistive limit.  In their
present form the simulations of Tao et al. fail to reach this limit
by a factor of roughly $25$.  Since we require that
$\eta$ must be no larger than $\nu$ this implies that
$\nu$ has to be lowered by a factor of roughly 25 if we keep
$\eta=\nu$.  A better strategy would be lower $\nu$ and leave
$\eta$ fixed, in which case the Reynolds number need only
go up by a factor $\sim5$ to reach the more realistic resistive
regime.

Another useful comparison can be made with the work of Nordlund
et al. (1992) who simulated a three dimensional dynamo in a
convective flow.  They found no field amplification for $\eta\ge\nu$,
but for $\nu=2\eta$ and $\nu=4\eta$ their simulations evolved towards
a unique stationary state with a kinetic energy approximately
$\sim 16$ the magnetic energy.  The Mach number of the flow was of
order $10^{-2}$ and the Reynolds number was $\sim 300$, based on the
thickness of the convective layer, which is roughly
$95$ by the definition we have used here.  We see from these parameters
that the flow was in the viscous regime, a bit above the dividing line
between moderate Reynolds numbers (where $V_A\approx V_T$) and large
Reynolds numbers.  According to the model given here their final
state should be a minimal energy state with roughly one flux tube
per turbulent cell and an energy ratio (from eq. (\ref{eq:z7}))
of
\begin{equation}
{V_A^2\over V_T^2}\approx {C_dw\pi^2\over 8} {\pi\over 300}\approx 0.039,
\end{equation}
where we have used $w\approx 2$ and $C_d\approx 1.5$.  This is
low by a factor of $\sim 1.6$, but given the more complicated nature of their
simulation this is still reasonable agreement with our model.
In fact, given that the simulation of Tao et al. indicated that we
underestimated the Reynolds number dividing the viscous regime
moderate and strong Reynolds numbers by a factor of $\sim 1.5$, and
that the ratio of magnetic energy to kinetic energy should drop
linearly with the inverse of the Reynolds number above this threshold,
the entire discrepancy may be due to this same point. This would
imply that the right hand side of eq. (\ref{eq:modvisc}) should
be multiplied by a factor of 1.5.
Nordlund et al. do not quote an average flux tube radius or
Alfv\'en velocity in the flux tubes.

These two simulations have Reynolds numbers which are
roughly similar, which makes it harder to draw firm conclusions from their
agreement with our model.  Moreover, the details of convective
turbulence are different from turbulence driven at large
scales by a random external force.  However, we note that these
simulations do show the expected
qualitative behavior, which is that for the minimal state of a
viscously dominated simulation the ratio of magnetic energy to
kinetic energy drops as the Reynolds number increases.

\clearpage
\begin{figure}
\plotone{fig1.eps}
\vskip -1.5cm
\caption{The fractional density depletion within flux tubes in the
Sun as a function of temperature, assuming equipartition between the
magnetic field and the local
turbulence.  When this is greater than one it is roughly the inverse
of the fraction of the flux tube volume occupied by the resistive skin.}
\end{figure}
\begin{figure}
\plotone{fig2.eps}
\caption{The magnetic flux in a typical solar flux tube as a function of
temperature, assuming equipartition between the magnetic field and the
local turbulence.}
\end{figure}
\begin{figure}
\vskip -1cm
\plotone{fig3.eps}
\vskip -2cm
\caption{The fraction of the local magnetic energy density that would
be present if the magnetic flux tubes didn't stretch as they rose, but
still rose at a systematic velocity which is a fixed fraction of the local
turbulent velocity.  This is the complement of the fraction of the total energy
in the magnetic field that is added in the form of small scale structures as
the flux tubes rise.}
\end{figure}
\end{document}